\documentclass[12pt]{article}
\usepackage{epsf,amssymb,amsmath,latexsym}
\title{Rotation numbers for Jacobi matrices with matrix entries}

\author{Hermann Schulz-Baldes
\\
\\
{\small Mathematisches Institut, Universit\"at
Erlangen-N\"urnberg, Germany}
}

\date{ }

\newtheorem{theo}{Theorem}

\newtheorem{proposi}{Proposition}
\newtheorem{lemma}{Lemma}

\newcommand{\CC}{{\mathbb C}}
\newcommand{\NN}{{\mathbb N}}
\newcommand{\RR}{{\mathbb R}}

\newcommand{\ZZ}{{\mathbb Z}}
\newcommand{\LM}{{\mathbb L}}
\newcommand{\KM}{{\mathbb K}}
\newcommand{\HM}{{\mathbb H}}
\newcommand{\DM}{{\mathbb D}}
\newcommand{\GM}{{\mathbb G}}
\newcommand{\UM}{{\mathbb U}}

\newcommand{\Pp}{{\cal P}}

\newcommand{\EE}{{\bf E}}

\newcommand{\Gg}{{\cal G}}

\newcommand{\Oo}{{\cal O}}
\newcommand{\Tr}{\mbox{\rm Tr}}
\newcommand{\Tt}{{\cal T}}
\newcommand{\Rr}{{\cal R}}
\newcommand{\Nn}{{\cal N}}
\newcommand{\Mm}{{\cal M}}
\newcommand{\Cc}{{\cal C}}
\newcommand{\Jj}{{\cal J}}
\newcommand{\Ii}{{\cal I}}

\newcommand{\Hh}{{\cal H}}

\newcommand{\one}{{\bf 1}}
\newcommand{\nul}{{\bf 0}}
\newcommand{\inv}{{\mbox{\rm\tiny inv}}}

\begin{document}

\maketitle

\begin{abstract}
A selfadjoined block tridiagonal matrix with positive definite blocks on the
off-diagonals is by definition a Jacobi matrix with matrix entries.
Transfer matrix techniques are extended
in order to develop a rotation number calculation for its eigenvalues.  
This is a matricial generalization of the oscillation theorem for the discrete
analogues of Sturm-Liouville operators. 
The three universality classes of time reversal invariance 
are dealt with by implementing the corresponding symmetries.
For Jacobi matrices with random matrix entries, this leads to
a formula for the integrated density of states which can
be calculated perturbatively in the coupling constant of the
randomness with an optimal control on the error terms.
\end{abstract}

\vspace{.5cm}

\section{Introduction}

This article is about matrices of the type
\begin{equation}
\label{eq-matrix}
H^N
\;=\;
\left(
\begin{array}{ccccccc}
V_1       & T_2  &        &        &         &        \\
T_2      & V_2    &  T_3  &        &         &        \\
            & T_3 & V_3    & \ddots &         &        \\
            &        & \ddots & \ddots & \ddots  &        \\
            &        &        & \ddots & V_{N-1} & T_N   \\
            &        &        &        & T_N  & V_N
\end{array}
\right)
\;,
\end{equation}
where $V_n=V_n^*$ are selfadjoint complex $L\times L$ matrices and 
$T_n$ are positive definite complex $L\times L$ matrices (Note added Feb.
2008: invertibility of $T_n$ is sufficient with minor modifications).
With the convention $T_1=\one$ and for a complex energy $E\in\CC$, 
introduce the transfer matrices
\begin{equation}
\label{eq-transfer}
\Tt_n^E
\;=\;
\left(
\begin{array}{cc}
(E\,{\bf 1}\,-\,V_n)\,T_n^{-1} & - T_n \\
T_n^{-1} & {\bf 0}
\end{array}
\right)
\;,
\qquad
n=1,\ldots,N
\;.
\end{equation}
Then set 
\begin{equation}
\label{eq-Uintro}
U_N^E
\;=\;
\left(
\begin{array}{c}
\one \\ \imath\,\one
\end{array}
\right)^*
\prod_{n=1}^N\,\Tt^E_n\,
\left(
\begin{array}{c}
\one \\ \nul
\end{array}
\right)
\;
\left(
\left(
\begin{array}{c}
\one \\ \imath\,\one
\end{array}
\right)^t
\prod_{n=1}^N\,\Tt^E_n\,
\left(
\begin{array}{c}
\one \\ \nul
\end{array}
\right)
\right)^{-1}
\;.
\end{equation}
%

\begin{theo} 
\label{theo-osci}
Let $E\in\RR$ and $N\geq 2$.

\noindent 
{\rm (i)} $U_N^E$ is well-defined, namely the appearing inverse exists.

\noindent {\rm (ii)} $U_N^E$ is
a unitary matrix which is real analytic in $E$. 

\noindent {\rm (iii)} The real
eigenphases $\theta^E_{N,l}$, $l=1,\ldots,L$, of $U_N^E$
can be chosen {\rm (}at level crossings{\rm )}  to be 

analytic in
$E$ and such that $\theta^E_{N,l}\to 0$ as
$E\to-\infty$ and $\theta^E_{N,l}\to 2\pi N$ as
$E\to\infty$. 

\noindent {\rm (iv)} $E$ is an
eigenvalue of $H^N$ of multiplicity $m$ 
if and only if $\theta^E_{N,l}=\pi\!\!\mod 2\pi$ for exactly $m$ 

of the indices
$l=1,\ldots,L$. 

\noindent {\rm (v)} The matrix $S^E_N=\frac{1}{\imath}\,
(U^E_N)^*\partial_EU^E_N$
is positive definite. Each $\theta^E_{N,l}$ is an increasing 
function of $E$.

\noindent {\rm (vi)} If $H^N$ is real, the unitary 
$U_N^E$ is symmetric and the positive matrix $S^E_N$ is real.

\noindent {\rm (vii)} Let $L$  be even and let 
\begin{equation}
\label{eq-selfdual}
I\;=\;
\left(
\begin{array}{cc}
\nul & - \one \\
\one & {\bf 0}
\end{array}
\right)
\;\in\;
\mbox{\rm Mat}(L\times L,\CC)
\end{equation}
\indent with $4$ blocks of size
$\frac{L}{2}\times \frac{L}{2}$. Suppose that $H^N$ is self-dual, namely the
entries are self-dual:
$$
I^*\,T_n^t\,I\;=\;T_n\;,
\qquad
I^*\,V_n^t\,I\;=\;V_n\;,
\qquad
n=1,\ldots,N\;.
$$

Then $U^E_N$ and $S^E_N$ are also
self-dual {\rm (}equivalently,
$IU_N^E$ and $IS_N^E$ are skew-symmetric{\rm )}.

\end{theo}

Items (i), (ii), (iv), (vi) and (vii) result directly from the mathematical
set-up, while the analyticity statements of 
items (ii) and (iii) are based on elementary 
analytic perturbation theory \cite{Kat}. The second part of (iii) follows from
a homotopy argument and item (v), even though a consequence of a
straight-forward calculation, is the main mathematical  insight.
It justifies the term
{\sl rotation numbers} for the eigenphases. In the strictly 
one-dimensional situation and for Sturm-Liouville operators instead of
Jacobi matrices, the theorem has been known for almost two cenuries 
as the rotation number calculation
or the Sturm-Liouville oscillation theorem \cite{Wei,JM}. For matricial
Sturm-Liouville operators, Bott \cite{Bot} has proven results related to the
above theorem. (The author learned of Bott's work once this article was
finished, and believes that the techniques presented below allow to
considerably simplify Bott's proof. A detailed treatment is under
preparation.) For related work on linear Hamiltonian system let us 
refer to the review \cite{FJN}.
The discrete one-dimensional case
and hence precisely the case $L=1$ of Theorem~\ref{theo-osci} is also 
well-known (see {\it e.g.} \cite{JSS} for a short proof). A rougher
result was proven by Arnold \cite{Arn2}. In the
one-dimensional situation the variable $\theta^E_{N,1}$ is called the Pr\"ufer
phase. Therefore one may refer to the eigenphases $\theta^E_{N,l}$ 
(or the unitaries $U^E_N$ themselves) 
also as multi-dimensional Pr\"ufer phases.
The two supplementary symmetries considered in items (vi) and (vii) 
correspond
to quantum-mechanical Hamiltonians $H^N$ with time-reversal invariance
describing systems with odd or even spin respectively \cite{Meh}.
This notion is empty in the one-dimensional situation where
time-reversal invariance follows automatically from self-adjointness.

\vspace{.2cm}

Crucial ingredient of the proof
is that \eqref{eq-Uintro}  for real energies actually stems 
from the M\"obius action of the symplectic transfer matrices
\eqref{eq-transfer} on the unitary
matrices (Theorem~\ref{theo-Usymdyn}), 
which in turn are diffeomorphic to the Lagrangian
Grassmannian via the stereographic projection
(Theorem~\ref{theo-realLag}). As a function of real energy, $U^E_N$ hence 
corresponds to a path of Lagrangian planes. If one defines a singular cycle in
the unitary group as the set of unitaries with eigenvalue $-1$, then the
intersections of the above path with this cycle turn out to be precisely at
the eigenvalues of $H^N$.

\vspace{.1cm}

One new perspective opened by the  Theorem~\ref{theo-osci}  concerns  
Jacobi matrices with random matrix entries, describing {\it
e.g.} finite volume approximations of the higher-dimensional Anderson 
model. In fact, the unitary, symmetric unitary and anti-symmetric unitary
matrices form precisely the state spaces of Dyson's circular   
ensembles. They are furnished with unique invariant measures 
(Haar measures on the corresponding symmetric spaces).
A good working hypothesis is hence that 
the random dynamical system induced by the action of the symplectic 
transfer matrices on the unitary matrices has an invariant measure
(in the sense of Furstenberg \cite{BL}) which is invariant under
the adequate action of the unitary group. 
This can only be true to lowest order in perturbation theory,
under a hypothesis on the coupling of the randomness (which has to be checked
for concrete models), and
on a set of lower dimensional unitary matrices corresponding to the
elliptic channels (in the sense of \cite{SB} and Section~\ref{sec-covariant}).
This then would
justify the random phase approximation or maximal entropy Ansatz 
(here as equidistribution of Lagrangian planes)
widely used in the physics community in the study of 
quasi-one-dimensional systems and
in order to establish a link between random models like the
Anderson Hamiltonian and invariant random matrix ensembles
({\it e.g.} \cite{Bee}). Furthermore, let us consider
$H^N$ describing a physical system on a $d$-dimensional cube, namely 
with $L=N^{d-1}$, and suppose $d$ sufficiently large.
Then a further working hypothesis is that the positive
matrices $\frac{1}{N}S^E_N$ are distributed according to the
Wishard ensemble of adequate symmetry 
(again on the elliptic subspaces and in the weak coupling
limit). If both working hypothesis turn out to hold and $U^E_N$ and $S_N^E$
are asymptotically independent, namely randomly rotated w.r.t. each
other (for large cubes), then 
Theorem~\ref{theo-osci} combined with a convolution argument shows that the
level statistics of $H^N$ is
asymptotically
given by the Wigner-Dyson statistics for quantum systems without or with
time-reversal invariance for odd or even spin, according to the symmetry of
$H^N$. Roughly, Theorem~\ref{theo-osci}
hence gives one possible way to materialize
the heuristics  given in the introduction to Chapter~9 of Mehta's
book \cite{Meh}. Numerical results supporting the above have been
obtained in collaboration with R. R\"omer. 


\vspace{.1cm}

As a first application of Theorem~\ref{theo-osci} and the techniques 
elaborated in its proof
we develop in Section~\ref{sec-covariant} 
the lowest order perturbation theory for the integrated
density of states (IDS) of a semi-infinite real 
Jacobi matrix with random matrix entries. More precisely, we
suppose that the entries
in \eqref{eq-matrix} are of the form $V_n=V(\one+\lambda\,v_n+\Oo(\lambda^2))$
and $T_n=T(\one+\lambda\,t_n+\Oo(\lambda^2))$ where $V,T,v_n,t_n$ are 
real symmetric
matrices and $T$ is positive definite. 
The $v_n,t_n$ are
 drawn independently and identically from a bounded ensemble
$(v_\sigma,t_\sigma)_{\sigma\in\Sigma}$ according to a given distribution
$\EE_\sigma$, and 
furthermore the dependence on the coupling constant 
$\lambda\geq 0$ is real analytic and the error
terms satisfy norm estimates. Associated to a random sequence 
$\omega=(v_n,t_n)_{n\geq 1}$ are random real Hamiltonians
$H^N(\omega,\lambda)$. The Anderson model on a strip is an example within this
class of models.
The number of eigenvalues of $H^N(\omega,\lambda)$
smaller than a given energy $E$ and per volume element $NL$ is a self-averaging
quantity in the limit $N\to\infty$ which converges to the IDS $\Nn_\lambda(E)$
(see Section~\ref{sec-covariant} for the formal definition). Let furthermore 
$\Nn_{\lambda,\sigma}(E)$ denote the IDS of the translation invariant
Hamiltonian with $\omega=(v_\sigma,t_\sigma)_{n\geq 1}$. 
Finally let $\Tt^E$ be the
transfer matrix defined as in \eqref{eq-transfer} from the unperturbed entries
$V$ and $T$. 

\vspace{-.1cm}

\begin{theo} 
\label{theo-perturb}
Suppose that $E\in\RR$ is such that $\Tt^E$ 
is diagonalizable and does not have anomalies, namely the rotation phases of
the elliptic channels are incommensurate 
{\rm (}cf. {\rm Section~\ref{sec-perturb}} for the precise hypothesis{\rm
  )}. Then 
$$
\Nn_{\lambda}(E)
\;=\;
\EE_\sigma
\bigl(
\Nn_{\lambda,\sigma}(E)
\bigr)
\;+\;
\Oo(\lambda^2)
\;.
$$
\end{theo}

The same result holds for random perturbations of arbitrary 
periodic operators. The fact
that $\Tt^E$ is not allowed to have any Jordan blocks means that $E$ is not an
internal band edge. Together with the anomalies they form a discrete set of
excluded energies.  
The main point of Theorem~\ref{theo-perturb} is
not the calculation of the leading order term in $\lambda$ (which is indeed
given by the most naive guess), but rather the control of the error
term which is uniform in $L$ as long as one stays uniformly bounded away from
anomalies and internal band edges. The estimates in
Section~\ref{sec-perturb} also show how the error bound diverges as one
approaches these energies. 
However, this part of the analysis is not optimized
and there is a definite need for refinement in order to be able to study the
thermodynamic limit of solid state physics models.
The error bound is nevertheless optimal in the sense that the $\Oo(\lambda^2)$
contribution on the l.h.s. does depend on further details of the model.
Similar as in the one-dimensional
situation ($L=1$), the IDS and the sum of the positive Lyapunov exponents are
imaginary and real boundary values of a single Herglotz function \cite{KS}. 
Hence we also
develop a perturbation theory for the sum of the Lyapunov exponents, with a
considerably better control on the error terms than in \cite{SB} where a
particular case has been treated.

\vspace{.1cm}

This work is organized as follows. The next section recollects the tools from
symplectic geometry used in the proof of Theorem~\ref{theo-osci}. In
particular, it is shown that the M\"obius action of the symplectic 
group on the unitary matrices is well-defined and furthermore some
formulas for the calculation of the intersection number (Maslov index) are
given. Section~\ref{sec-JMME}  provides the proof of
Theorem~\ref{theo-osci}
and then gives some supplementary results on Jacobi
matrices with matrix entries and their spectra. Section~\ref{sec-covariant}
contains the definition of the 
IDS for Jacobi matrices with random matrix entries
and a formula for the associated averaged Lyapunov exponent. Then the proof of
Theorem~\ref{theo-perturb} is given.

\vspace{.1cm}

\noindent {\bf Acknowledgment:} The author thanks Hajo Leschke, Demetris
Pliakis and Robert Schrader for discussions on the matters of the paper.
This work was supported by the DFG.


\section{Symplectic artillery}

Apart from the propositions in Section~\ref{sec-Arnold}
which may be strictly speaking new,
this section is probably known to the experts in symplectic
geometry. But there does not seem to be reference with a treatment as compact
and unified as the present one.
The author's references were \cite{Hua,Sie,Arn,CL,KSc,Arn3} 
and he hereby excuses for not
citing all the interesting works that he does not know of. The reader is
warned that the complex Lagrangian planes and the 
complex symplectic group are defined with the adjoint rather
than the transpose. This differs from standard references, but hopefully the
reader will agree that it is natural in
the present context.

\vspace{.2cm}

Let us introduce some notations.
The following $2L\times 2L$ matrices (matrices of this size
are denoted by mathcal symbols in this work) are composed by 4 blocks of size
$L\times L$:
$$
{\cal J}
\;=\;
\left(
\begin{array}{cc}
{\bf 0} & -{\bf 1} \\
{\bf 1} & {\bf 0}
\end{array}
\right)
\;,
\quad
{\cal G}
\;=\;
\left(
\begin{array}{cc}
{\bf 1} & {\bf 0} \\
{\bf 0} & -{\bf 1}
\end{array}
\right)
\;,
\quad
\Cc
\;=\;\frac{1}{\sqrt{2}}\;
\left(
\begin{array}{cc} {\bf 1} & -\,\imath\,{\bf 1} \\  
{\bf 1} & \imath\,{\bf 1} \end{array}
\right)
\;,
\quad
\Ii
\;=\;
\left(
\begin{array}{cc}
\nul & - I \\
I & {\bf 0}
\end{array}
\right)
\;,
$$
where in the last equation $I$ is given by \eqref{eq-selfdual} and hence $L$
is supposed to be even. $\Jj$ is called the symplectic form, $\Cc$ the Cayley
transform and $\Ii$ the self-duality transform. The following identities will
be useful:
\begin{equation}
\label{eq-Cayley}
\Cc\, \Jj\,\Cc^*
\;=\;
\frac{1}{\imath}\;\Gg\;,
\qquad
\overline{\Cc}\, \Jj\,\Cc^*
\;=\;
\frac{1}{\imath}\;\Jj\;,
\qquad
\overline{\Cc}\,\Ii\,\Cc^*\;=\;
\frac{1}{\imath}\;\Ii\;.
\end{equation}

In order to deal with the symmetry of Theorem~\ref{theo-osci}(vii), hence $L$
even, some further notations are convenient. A matrix $A\in \mbox{Mat}(L\times
L,\CC)$ is call self-dual if $I^*A^tI=A$, and it is called self-conjugate if
$I^*\overline{A}I=A$. As already indicated in Theorem~\ref{theo-osci}(vii),
self-duality is closely linked to skew-symmetry, namely
$A$ is self-dual if and only if $(IA)^t=-IA$. The sets of skew-symmetric and
self-dual matrices are denoted by Skew$(L,\CC)$ and Self$(L,\CC)$
respectively. Moreover,
for selfadjoint matrices $A^*=A$ the notions of self-duality and
self-conjugacy coincide.

\subsection{Lagrangian Grassmannian}

The set of $L$-dimensional subspaces of the complex vector space
$\CC^{2L}$ is denoted by $\GM^\CC_L$.
The vectors of a basis of such a plane form the column vectors of a 
$2L\times L$ matrix $\Phi$ which has rank $L$.
Of course,
a plane does not depend on the choice of the basis (and hence the
explicit form of $\Phi$). Consider  
the relation: $\Phi \sim\Psi$ $\Leftrightarrow$ there exists
$c\in\,$Gl$(L,\CC)$ with $\Phi=\Psi c$. The Grassmannian $\GM_L^\CC$ 
is then the set of equivalence classes  w.r.t. $\sim$: 
$$
\GM_L^\CC
\;=\;
\left\{
[\Phi]_{\sim} \;\left|\;
\Phi\in\mbox{Mat}(2L\times L,\CC)\;,\;\;\mbox{rank}(\Phi)=L\;
\right.\right\}
\;.
$$

A plane is called (complex hermitian) Lagrangian if $\Phi^*\Jj\Phi={\bf 0}$. 
Here $A^*=\overline{A}^t$ denotes transpose of 
the complex conjugate of a matrix $A$. If
$\Phi=\left(\begin{array}{c} a \\  b \end{array}\right)$ where $a$ and $b$ are
complex $L\times L$ matrices,
the latter condition means that $a^*b=b^*a$ is selfadjoint.
The (complex hermitian) Lagrangian Grassmannian $\LM_L^\CC$ is the
set of Lagrangian planes: 
\begin{equation}
\label{eq-complag}
\LM_L^\CC
\;=\;
\left\{
[\Phi]_{\sim} \;\left|\;
\Phi\in\mbox{Mat}({2L\times L},\CC)\;,\;\;\mbox{rank}(\Phi)=L\;,
\;\;\Phi^*\Jj\Phi={\bf 0}\;
\right.\right\}
\;.
\end{equation}
This is a real analytic manifold.
It contains the submanifold $\LM_L^\RR$ of real Lagrangian planes:
\begin{equation}
\label{eq-reallag}
\LM_L^\RR
\;=\;
\left\{
[\Phi]_{\sim} \;\left|\;
\Phi\in\mbox{Mat}({2L\times L},\CC)\;,\;\;\mbox{rank}(\Phi)=L\;,
\;\;\Phi^*\Jj\Phi={\bf 0}\;,\;\;\Phi^t\Jj\Phi={\bf 0}\;
\right.\right\}
\;.
\end{equation}
Hence $\LM_L^\RR$ is a subset of $\LM_L^\CC$ characterized by a
supplementary symmetry. That this coincides with the usual definition of the
real Lagrangian Grassmannian is stated in Theorem~\ref{theo-equidef}
below. If $L$ is even, then $\LM_L^\CC$ contains another submanifold
characterized by another symmetry:
\begin{equation}
\label{eq-quatlag}
\LM_L^\HM
\;=\;
\left\{
[\Phi]_{\sim} \;\left|\;
\Phi\in\mbox{Mat}({2L\times L},\CC)\;,\;\;\mbox{rank}(\Phi)=L\;,
\;\;\Phi^*\Jj\Phi={\bf 0}\;,\;\;\Phi^t\Ii\Phi={\bf 0}\;
\right.\right\}
\;.
\end{equation}
The notation $\LM_L^\HM$ appealing to the
quaterions is justified by the following theorem, in which $\HM^L$ is
considered as a vector space over $\CC$. Let $A^{*_\HM}$
denote the transpose and quaternion conjugate (inversion of sign
of all three imaginary parts) of a matrix $A$ with quaternion entries.
Similarly, $A^{*_\RR}=A^t$ for a matrix with real entries.

\begin{theo}
\label{theo-equidef}
One has, with equality in the sense of diffeomorphic real analytic manifolds,
$$
\LM_L^\RR
\;=\;
\left\{
[\Phi]_{\sim} \;\left|\;
\Phi\in\mbox{\rm Mat}({2L\times L},\RR)\;,\;\;\mbox{\rm rank}(\Phi)=L\;,
\;\;\Phi^{*_\RR}\Jj\Phi={\bf 0}\;
\right.\right\}
\;,
$$
and
$$
\LM_L^\HM
\;=\;
\left\{
[\Phi]_{\sim} \;\left|\;
\Phi\in\mbox{\rm Mat}({L\times L},\HM)\;,\;\;\mbox{\rm rank}(\Phi)=L\;,
\;\;\Phi^{*_\HM} I\Phi={\bf 0}\;
\right.\right\}
\;.
$$
\end{theo}

\vspace{.2cm}

The proof is postponed to the next section.

\subsection{Stereographic projection}
\label{sec-realLag}

Bott \cite{Bot} showed that the complex Lagrangian Grassmannian
$\LM^\CC_L$ is homeomorphic to the unitary group U$(L)$, a fact that was
rediscovered in \cite{KSc,Arn3}.
Arnold \cite{Arn} used the fact that $\LM^\RR_L$ can be identified with
U$(L)/\,$O$(L)$. Indeed,  
a real Lagrangian plane can always be spanned by an orthonormal
system, that is, be represented by
$\Phi$ satisfying $\Phi^*\Phi=\one$. This induces
$(\Phi,\Jj\Phi)
\in\,$SP$(2L,\RR)\cap\,$O$(2L)\cong\,$U$(L)$. Of course,
various orthonormal systems obtained by orthogonal basis changes within the
plane span the same Lagrangian  plane. Hence
$\LM^\RR_L\cong\,$U$(L)/\,$O$(L)$. Moreover, the
symmetric space U$(L)/\,$O$(L)$ can be identified with the unitary symmetric
matrices by sending a right equivalence class
$A\,\mbox{O}(L)\in\;$U$(L)/\,$O$(L)$ to $AA^t$.
As these facts will be crucial later on, let us give a detailed proof and some
explicit formulas.

\vspace{.2cm}

The stereographic projection $\pi$ is defined on the subset 
$$
\GM_L^\inv
\;=\;
\left\{
[\Phi]_{\sim}\in \GM_L\;\left|\;
\bigl({\bf 0} \, \one\bigr)\,\Phi
\,\in\,\mbox{GL}(L,\CC)
\right.\right\}
\;,
$$
by
$$
\pi
([\Phi]_\sim)
\;=\;
\bigl(\one\, {\bf 0}\bigr)\,\Phi
\; 
\bigl(\,\bigl({\bf 0} \, \one\bigr)\,\Phi\bigr)^{-1}
\;=\;
a\,b^{-1}
\;,
\qquad
\Phi\;=\;
\left(
\begin{array}{c} a \\  b \end{array}
\right)
\;.
$$
One readily checks that $\pi([\Phi]_\sim)$ is 
independent of the representative. 
If $[\Phi]_\sim\in\LM^\CC_L\cap \GM_L^\inv$, then 
$\pi([\Phi]_\sim)$ is selfadjoint.
The fact that $\pi$ is not defined on
all of $\LM_L^\CC$ is an unpleasant feature that can be circumvented by use
of $\Pi$ defined by
$$
\Pi([\Phi]_\sim)
\;=\;
\pi
([\Cc\,\Phi]_\sim)
\;,
\qquad
\mbox{\rm if }\;[\Cc\,\Phi]_\sim\in\GM_L^\inv
\;.
$$

\begin{theo}
\label{theo-realLag}
{\rm (i)} The map $\Pi: \LM_L^\CC\to\, ${\rm U}$(L)$ is a real analytic
diffeomorphism. 

\vspace{.1cm}

\noindent {\rm (ii)}
The map $\Pi: \LM_L^\RR\to\, ${\rm U}$(L)\,\cap\;${\rm Sym}$(L,\CC)$
is a real analytic diffeomorphism.

\vspace{.1cm}

\noindent {\rm (iii)} Let $L$ be even.
The map $\Pi: \LM_L^\HM\to\, ${\rm U}$(L)\,\cap\;${\rm Self}$(L,\CC)$
is a real analytic diffeomorphism.
\end{theo}

\noindent {\bf Proof.} Let $\Phi=\left(\begin{array}{c} a \\  b
\end{array}\right)$ where $a$ and $b$ are $L\times L$ matrices satisfying
$a^*b=b^*a$. One has
\begin{eqnarray}
L 
& = &
\mbox{rank}(\Phi)
\;=\;
\mbox{rank}(\Phi^*\Phi)
\;=\;
\mbox{rank}(a^*a+b^*b)
\nonumber
\\
\label{eq-rank}
& & 
\\
& = &
\mbox{rank}\bigl((a+\imath \,b)^*(a+\imath \,b) \bigr)
\;=\;
\mbox{rank}(a+\imath \,b)
\;=\;
\mbox{rank}(a-\imath \,b)
\;.
\nonumber
\end{eqnarray}
It follows that $[\Cc\Phi]_\sim\in \GM^\inv_L$ so that it is in the domain of
the stereographic projection $\pi$ and hence $\Pi$ is
well-defined. Next let us show 
that the image is unitary. It follows from \eqref{eq-rank} and a short
calculation (or alternatively the first identity in \eqref{eq-Cayley}) that
$$
\Cc\;\LM_L^\CC
\;=\;
\left.\left\{
\Bigl[ \Bigl(\begin{array}{c} a \\  b
\end{array}\Bigr)\Bigr]_{\sim} \;\right|\;
a,b\in\mbox{GL}({L},\CC)\;,\;\;a^*a=b^*b\;
\right\}
\;.
$$
Hence if $[\Cc\Phi]_\sim=\Bigl[ \Bigl(\begin{array}{c} a \\  b
\end{array}\Bigr)\Bigr]_{\sim}\in\Cc\,\LM_L^\CC$, one has
$$
\Pi([\Phi]_\sim)^*\Pi([\Phi]_\sim)
\;=\;
(b^*)^{-1}a^*a\,b^{-1}
\;=\;\one
\;.
$$
Moreover, $\Pi$ is continuous. 
One can directly check that the inverse of $\Pi$ is given by
\begin{equation}
\label{eq-Piinv}
\Pi^{-1}(U)
\;=\;
\left[
\left(
\begin{array}{c} \frac{1}{2}(U+\one) \\  \frac{\imath}{2}(U-\one) 
\end{array}\right)
\right]_\sim
\;.
\end{equation}
As this is moreover real analytic, this proves (i).

For the case (ii) of the real Lagrangian Grassmannian, the second identity of
\eqref{eq-Cayley} implies that the supplementary symmetry in 
\eqref{eq-reallag} leads to
$$
\Cc\;\LM_L^\RR
\;=\;
\left.\left\{
\Bigl[ \Bigl(\begin{array}{c} a \\  b
\end{array}\Bigr)\Bigr]_{\sim} \;\right|\;
a,b\in\mbox{GL}({L},\CC)\;,\;\;a^*a=b^*b\;,
\;\;\;(ab^{-1})^t=ab^{-1}\;
\right\}
\;.
$$
This implies that 
$\Pi([\Phi]_\sim)$ is symmetric for $[\Phi]_\sim\in\LM_L^\RR$.
Moreover, if $U$ in \eqref{eq-Piinv} is symmetric, then the last identity in 
\eqref{eq-reallag} holds. 
Again $\Pi$ is continuous, and $\Pi^{-1}$ real analytic,
so that the proof of (ii) is completed.

For case (iii), the third identity of
\eqref{eq-Cayley} implies that the supplementary symmetry in 
\eqref{eq-quatlag} gives
$$
\Cc\;\LM_L^\HM
\;=\;
\left.\left\{
\Bigl[ \Bigl(\begin{array}{c} a \\  b
\end{array}\Bigr)\Bigr]_{\sim} \;\right|\;
a,b\in\mbox{GL}({L},\CC)\;,\;\;a^*a=b^*b\;,
\;\;\;I^*(ab^{-1})^tI=ab^{-1}\;
\right\}
\;.
$$
This implies that 
$I\Pi([\Phi]_\sim)$ is skew-symmetric for $[\Phi]_\sim\in\LM_L^\HM$.
Again, if $IU$ in \eqref{eq-Piinv} is skew-symmetric, 
then the last identity in \eqref{eq-quatlag} holds, completing the proof. 
\hfill $\Box$

\vspace{.2cm}

\noindent {\bf Proof} of Theorem~\ref{theo-equidef}. Let $\hat{\LM}^\RR_L$ and 
$\hat{\LM}^\HM_L$ denote the real analytic manifolds on the r.h.s. of the two
equations in Theorem~\ref{theo-equidef}. Let us first show 
${\LM}^\RR_L=\hat{\LM}^\RR_L$. The inclusion $\hat{\LM}^\RR_L
\subset{\LM}^\RR_L$ is obvious because for a real representative $\Phi$ the
two conditions in \eqref{eq-reallag} coincide. Moreover, this inclusion is
continuous. Due to Theorem~\ref{theo-realLag}(ii) it is sufficient to
show that $\Pi^{-1}:\,${\rm U}
$(L)\,\cap\;${\rm Sym}$(L,\CC)\to \hat{\LM}^\RR_L$, namely that one can choose
a real representative in \eqref{eq-Piinv}. 
Let us use the fact that every symmetric unitary $U$ can
be diagonalized by an orthogonal matrix $M\in\,$O$(L)$, namely $U=M^tDM$ where
$D=\mbox{diag}(e^{\imath\theta_1},\ldots, e^{\imath\theta_L})$ with
$\theta_l\in [0,2\pi)$. Now let 
$D^{\frac{1}{2}}=
\mbox{diag}(e^{\imath\theta_1/2},\ldots, e^{\imath\theta_L/2})$
be calculated with  the first branch of the square root and choose
$S=\mbox{diag}(\sigma_1,\ldots, \sigma_L)\in\,$O$(L)$ 
with $\sigma_l\in\{-1,1\}$
such that the phase of $\sigma_le^{\imath\theta_l/2}$ is in $[0,\pi)$. Then
let us introduce the unitary 
$V=M^tD^{\frac{1}{2}}S$. One has $U=VV^t$. Furthermore set
$a=\Re e(V)$ and $b=-\Im
m(V)$ and then $\Pi^{-1}(U)=[\Phi]_\sim$ with 
$\Phi=\left(\begin{array}{c} a \\  b \end{array}\right)$.
Indeed $\Pi^{-1}$ is the inverse of $\Pi$:
$$
\Pi([\Phi]_\sim)
\;=\;
\pi\left(\left[\left(\begin{array}{c} V \\ \overline{V} \end{array}
\right)\right]_\sim\right)
\;=\;
VV^t
\;=\;U
\;.
$$

This construction of $\Pi^{-1}$ was done with a bit more care than needed, but
it allows to show directly that $\Pi^{-1}$ is locally real analytic.
Let $E\mapsto U(E)$ be a real analytic path of symmetric
unitaries. Then analytic perturbation theory \cite[Theorem II.1.10]{Kat} 
shows that the
diagonalization $U(E)=M(E)^tD(E)M(E)$ can be done with
analytic $M(E)$ and $D(E)$. Furthermore  
$E\mapsto D(E)^{\frac{1}{2}}S(E) \in
\mbox{diag}(\RR/\pi\ZZ,\ldots,\RR/\pi\ZZ)$ 
with $S(E)$ defined as above is
also analytic because $\theta\in \RR/2\pi\ZZ\mapsto 
\frac{\theta}{2}\in \RR/\pi\ZZ$ is analytic. Thus $V(E)=
M(E)^tD^{\frac{1}{2}}(E)S(E)$ is analytic
and therefore also $\Pi^{-1}$. 

The proof of ${\LM}^\HM_L=\hat{\LM}^\HM_L$ is just an adapted 
version of the usual rewriting of symplectic structures (here the symmetry
induced by $I$ and $\Ii$) in terms of quaternions. Let the basis of $\HM$ (as
real vector space) be $1$ and the imaginary units $\imath,j,k$ satisfying
Hamilton's equations $\imath^2=j^2=k^2=\imath jk=-1$. Then $\CC$ is identified
with the span of $1$ and $\imath$. Now let us introduce
$$
\Upsilon
\;=\;
\left(\begin{array}{cccc} \one & j\,\one & \nul & \nul \\ 
\nul & \nul &  \one & j\,\one \end{array}
\right)
\;\in\;
\mbox{Mat}(L\times 2L,\HM)
\;,
$$
where all blocks are of size $\frac{L}{2}\times \frac{L}{2}$. One readily
verifies the matrix identity 
\begin{equation}
\label{eq-quattrans}
\Upsilon^{*_\HM}\,I\,\Upsilon
\;=\;
\Jj\;-\;j\,\Ii
\;.
\end{equation}
Hence one obtains a map 
$\Upsilon:\,\mbox{Mat}(2L\times L,\CC)\to\mbox{Mat}(L\times L,\HM)$  by matrix
multiplication with $\Upsilon$ (from the left) which
induces a map on the (complex) Grassmannians of right equivalence
classes. Thus due to \eqref{eq-quattrans} we have exhibited an analytic map
$\Upsilon:\hat{\LM}^\HM_L\to{\LM}^\HM_L$. 
\hfill $\Box$

\subsection{Symplectic group and Lorentz group}

Let $\KM$ be one of the fields $\RR$, $\CC$ and $\HM$ and let $L$ 
be even if $\KM=\HM$. 
The symplectic group SP$(2L,\KM)$ is by definition the set of complex 
$2L\times 2L$ matrices conserving the Lagrangian structure in
\eqref{eq-complag}, \eqref{eq-reallag} and \eqref{eq-quatlag} respectively,
{\it e.g.} 
$$
\mbox{SP}(2L,\RR)
\,=\,
\left\{
\left.
\Tt
\,\in\,\mbox{Mat}({2L},\CC)
\;\right|
\;\Tt^*\Jj\Tt=\Jj\;,\;\;\;\Tt^t\Jj\Tt=\Jj
\,
\right\}
\;.
$$
One verifies that $\Tt\in\,$ SP$(2L,\KM)$ 
if and only if $\Tt^*\in\,$ SP$(2L,\KM)$. All
symplectic matrices have a unit determinant.
Using the Jordan form, it can be proven that SP$(2L,\KM)$ is 
arc-wise connected. Theorem~\ref{theo-equidef} implies respectively
the identity and isomorphism
(direct algebraic proofs can be written out as well)
$$
\mbox{SP}(2L,\RR)
\,=\,
\left\{
\left.
\Tt
\,\in\,\mbox{Mat}({2L},\RR)
\;\right|
\;\Tt^{*_\RR}\Jj\Tt=\Jj
\,
\right\}
\;,
$$
and
$$
\mbox{SP}(2L,\HM)
\,\cong\,
\left\{
\left.
T
\,\in\,\mbox{Mat}({L},\HM)
\;\right|
\;T^{*_\HM}IT=I
\,
\right\}
\;.
$$
More explicit formulas are given in the next algebraic lemma.

\begin{lemma}
\label{lem-sympexpl} 
The complex symplectic group is given by
$$
\mbox{\rm SP}(2L,\CC)
\,=\,
\left\{
\left(
\left.
\begin{array}{cc} A & B \\  C & D \end{array}
\right)
\,\in\,\mbox{\rm Mat}({2L},\CC)
\;\right|
\;
A^*C=C^*A\;,\;\;A^*D-C^*B={\bf 1}\;,\;\;B^*D=D^*B\,
\right\}
\;.
$$
In this representation, elements of $\mbox{\rm SP}(2L,\RR)$ and 
$\mbox{\rm SP}(2L,\HM)$ are characterized by having respectively real and
self-conjugate entries $A,B,C,D$.
\end{lemma}

As already became apparent in the proof of Theorems~\ref{theo-equidef} and
\ref{theo-realLag}, it is convenient to use the Cayley transform. The
generalized Lorentz groups are introduced by
$$
\mbox{U}(L,L,\KM)
\;=\;
\Cc\,\mbox{SP}(2L,\KM)\,\Cc^*
\;.
$$
From the identities \eqref{eq-Cayley} one can read off alternative 
definitions, {\it e.g.} 
$$
\mbox{U}(L,L,\RR)
\;=\;
\left\{\left.
\Tt\,\in\,\mbox{Mat}(2L\times 2L,\CC)
\;\right|
\;
\Tt^*\,\Gg\,\Tt\,=\,\Gg
\,,\;\;
\;
\Tt^t\,\Jj\,\Tt\,=\,\Jj
\,
\right\}
\;.
$$
Let us provide again more explicit expressions.

\begin{lemma}
\label{lem-Cayley} 
One has 
\begin{eqnarray}
\mbox{\rm U}(L,L,\CC)
\!\!\!
& = &
\!\!\!
\left\{
\left(
\left.
\begin{array}{cc} A & B \\  C & D \end{array}
\right)
\in\,\mbox{\rm Mat}({2L},\CC)
\,\right|
\,
A^*A-C^*C=\one\,,\,D^*D-B^*B=\one\,,\,
A^*B=C^*D\,
\right\}
\nonumber
\\
& = &
\!\!\!
\left\{
\left(
\left.
\begin{array}{cc} A & B \\  C & D \end{array}
\right)
\in\,\mbox{\rm Mat}({2L},\CC)
\,\right|
\,
AA^*-BB^*=\one\,,\,DD^*-CC^*=\one\,,\,
AC^*=BD^*\,
\right\}
.
\nonumber
\end{eqnarray}
Furthermore, in that representation, $A$ and $D$ are invertible and 
 $\|A^{-1} B\|<1$ and $\|D^{-1} C\|<1$. For $\Tt\in \mbox{\rm
 U}(L,L,\RR)$ one, moreover, has $C=\overline{B}$ and $D=\overline{A}$.
For $\Tt\in \mbox{\rm U}(L,L,\HM)$ the entries satisfy
$C=I^*\overline{B}I$ and $D=I^*\overline{A}I$.
\end{lemma}

\noindent {\bf Proof.} The first relations are equivalent to
$\Tt^*\Gg\Tt=\Gg$, the second ones then
follow from the fact that $\Tt^*\in
\mbox{\rm U}(L,L,\CC)$ for $\Tt\in \mbox{\rm U}(L,L,\CC)$. 
The fact that $A$ is invertible follows from
$AA^*\geq \one$. Furthermore 
$AA^*-BB^*={\bf 1}$ implies that $A^{-1}B(A^{-1}B)^*
=\one - A^{-1}(A^{-1})^*<\one$, so that $\|A^{-1} B\|<1$. The same argument
applies to $D$ and $D^{-1}C$. The last two statements
can be checked by a short calculation.
\hfill $\Box$

\subsection{Upper half-planes and Cartan's classical domains}
\label{sec-disc}
 
The upper half-plane and unit disc are defined by
$$
\UM_L^\CC
\,=\,
\left\{
Z\in\mbox{Mat}(L\times L,\CC)
\;\left|\;
\imath(Z^*-Z) >0\;
\right\}
\right.
\,,
\quad
\DM_L^\CC
\,=\,
\left\{
U\in\mbox{Mat}(L\times L,\CC)
\;\left|\;U^*U<{\bf 1}
\;
\right\}
\right.
\,,
$$
where $Y>0$ means that $Y$ is positive definite. Furthermore let us
introduce the following subsets (here $L$ does not need to be even for
$\KM=\HM$):
$$
\UM_L^\RR
\;=\;
\UM_L^\CC\,\cap\,\mbox{Sym}(L,\CC)
\;,
\qquad
\UM_L^\HM
\;=\;
\UM_L^\CC\,\cap\,\mbox{Self}(L,\CC)
\;,
$$
and
$$
\DM_L^\RR
\;=\;
\DM_L^\CC\,\cap\,\mbox{Sym}(L,\CC)
\;,
\qquad
\DM_L^\HM
\;=\;
\DM_L^\CC\,\cap\,\mbox{Self}(L,\CC)
\;.
$$
Let us note that $I\DM_L^\HM=
\DM_L^\CC\,\cap\,\mbox{Skew}(L,\CC)$.
The sets
$\DM^\CC_L$, $\DM_L^\RR$ and $I\DM_L^\HM$
are called Cartan's first, second and third classical
domain \cite{Hua}. Furthermore $\UM_L^\RR$ and $\DM_L^\RR$ are also called
the Siegel upper half-plane and the Siegel disc \cite{Sie}. The Cayley
transform maps (via M\"obius transformation) the upper half-planes
bijectively to the generalized unit discs, as shown next.

\begin{proposi}
\label{prop-dischalfplane}
The formulas
$$
U\;=\;(Z\,-\,\imath\,\one)(Z\,+\,\imath\,\one)^{-1}
\;,
\qquad
Z\;=\;\imath\;(\one\,+\,U)(\one\,-\,U)^{-1}
\;,
$$
establish an analytic diffeomorphism 
from $\UM_L^\KM$ onto $\DM_L^\KM$ for $\KM=\CC,\RR,\HM$.
\end{proposi}

\noindent {\bf Proof.} ({\it cf.} \cite{Sie}; 
reproduced for the convenience of the reader.) If
$v\in\,\mbox{ker}(Z+\imath\one)$, then $\imath v=-Zv$ so that
$0\leq \langle v|\imath(Z^*-Z)|v\rangle=-2\,\langle v|v\rangle$ which implies
$v=0$. Hence $Z+\imath\one$ is invertible and the first formula is 
well-defined. Similarly one checks the invertibility of $\one-U$. To verify
that one is the inverse of the other is a matter of calculation. Moreover, both
formulas preserve the symmetry and self-duality of the matrices involved.
\hfill $\Box$

\vspace{.2cm}

The boundary $\partial\UM_L^\CC$ of $\UM_L^\CC$ 
is a stratified space given as the
union of strata $\partial_l \UM^\CC_L$, $l=1,\ldots,L$, 
where $\partial_l \UM_L^\CC$  is the
set of matrices $Z$ for which $\imath (Z^*-Z)\geq 0$ is of rank
$L-l$. The maximal boundary is $\partial_L \UM_L^\CC$ are the selfadjoint
matrices. Furthermore 
$\partial\UM_L^\RR=\partial\UM^\CC_L\,\cap\,$Sym$(L,\CC)$ with strata 
$\partial_l\UM_L^\RR=\partial_l\UM_L^\CC\,\cap\,$Sym$(L,\CC)$. Corresponding 
formulas hold for $\KM=\HM$.

\vspace{.2cm}

Similarly the boundary $\partial\DM_L^\CC$ is a stratified space with strata 
$\partial_l\DM_L^\CC$, $l=1,\ldots,L$,
of matrices $U$ for which $U^*U\leq \one$ and 
rank$(\one-U^*U)=L-l$. For
$\KM=\RR,\HM$  one defines in the same way  the stratified boundaries 
$\partial\DM_L^\KM=\cup_{l=1}^L\partial_l\DM_L^\KM$. Of particular importance
will be the maximal boundaries ($L$ even for $\KM=\HM$):
$$
\partial_L\DM_L^\CC
\;=\;
\mbox{U}(L)
\;,
\qquad
\partial_L\DM_L^\RR
\;=\;
\mbox{U}(L)\cap\,\mbox{Sym}(L,\CC)
\;,
\qquad
\partial_L\DM_L^\HM
\;=\;
\mbox{U}(L)\cap\,\mbox{Self}(L,\CC)
\;.
$$
By Theorem~\ref{theo-realLag} the maximal boundary $\partial_L\DM_L^\KM$
is hence identified with the Lagrangian Grassmannian $\LM_L^\KM$.
Let us also note that the Cayley transformation of 
Proposition~\ref{prop-dischalfplane} has singularities on the boundaries and
mixes the strata. In particular, $\partial_L\UM_L^\KM$ is not mapped to
$\partial_L\DM_L^\KM$.

\subsection{M\"obius action}
\label{sec-Moebius}

The M\"obius transformation (also called
canonical transformation or fractional transformation) is defined by
\begin{equation}
\label{eq-moebius}
\Tt\cdot Z
\;=\;
(AZ+B)\,(CZ+D)^{-1}
\,,
\qquad
\Tt
\,=\,
\left(
\begin{array}{cc} A & B \\  C & D \end{array}
\right)
\;\in\;\mbox{GL}(2L,\CC)
\,,
\;\;Z\in\mbox{Mat}(L\times L,\CC)\,,
\end{equation}
whenever the appearing inverse exists.
This action implements the matrix multiplication, namely
\begin{equation}
\label{eq-implement}
\pi\bigl(\Tt([\Phi]_{\sim})\bigr)
\;=\;
\Tt\cdot\pi\bigl([\Phi]_{\sim}\bigr)
\;,
\qquad
\mbox{if }\;[\Phi]_{\sim}\in\GM_L^\inv
\mbox{ and }\;[\Tt\Phi]_{\sim}\in\GM_L^\inv
\;.
\end{equation}
Indeed, let $\Phi=\left(\begin{array}{c} a \\  b \end{array}\right)$.
Then $[\Tt\Phi]_{\sim}\in\GM_L^\inv$ implies that $Ca+Db$ is invertible. As
$b$ is invertible (because $[\Phi]_{\sim}\in\GM_L^\inv$), it follows that 
$Cab^{-1}+D=C\pi([\Phi]_\sim)+D$ 
is invertible so that the M\"obius transformation 
$\Tt\cdot\pi\bigl([\Phi]_{\sim}\bigr)$ is well-defined.
The conditions in \eqref{eq-implement} are automatically satisfied in the
situation of the
following proposition. The proof of item (ii) is contained in the proof of
Theorem~\ref{theo-Usymdyn} below; item (i) then follows directly from (ii)
due to  Proposition~\ref{prop-dischalfplane}.

\begin{proposi}
\label{prop-halfspacedyn} {\rm \cite{Hua,Sie}} Let $\KM=\CC,\RR,\HM$ and let 
$L$ be even for $\KM=\HM$.

\vspace{.1cm}

\noindent {\rm (i)} 
$\mbox{\rm SP}(2L,\KM)$ acts on $\UM_L^\KM$ by M\"obius transformation.

\vspace{.1cm}

\noindent {\rm (ii)} $\mbox{\rm U}(L,L,\KM)$ acts on 
$\DM_L^\KM$ by M\"obius transformation.
\end{proposi}

\vspace{.2cm}

The following proposition states that the action of
Proposition~\ref{prop-halfspacedyn}(ii) extends to the stratified
boundary of $\DM^\KM_L$. Moreover, the action on the
maximal boundary 
$\partial_L\DM^\KM_L$ implements 
the natural action of the
symplectic group on the Lagrangian Grassmannian. Due to
singularities it is not possible to extend the action of 
Proposition~\ref{prop-halfspacedyn}(i) 
to any stratum of the boundary of $\UM_L^\KM$.

\begin{theo}
\label{theo-Usymdyn}
Let $\KM=\CC,\RR,\HM$ and $L$ even if $\KM=\HM$ and $l=1,\ldots,L$. 
The Lorentz group $\mbox{\rm U}(L,L,\KM)$ 
acts on $\partial_l\DM^\KM_L$ by
M\"obius transformation. For the case $l=L$ of the maximal boundary, 
this action implements the action of 
$\mbox{\rm SP}(2L,\KM)$ on the Lagrangian Grassmannian $\LM^\KM_L$:
$$
\Pi\bigl([\Tt\Phi]_{\sim}\bigr)
\;=\;
\Cc\,\Tt\,\Cc^*\cdot\Pi\bigl([\Phi]_{\sim}\bigr)
\;,
\qquad
[\Phi]_{\sim}\in\LM_L^\KM\;,\;\;
\;\Tt\in\;\mbox{\rm SP}(2L,\KM)
\;.
$$
\end{theo}

\noindent {\bf Proof.} 
One has to show that
for $U\in \partial_l\DM_L^\KM$ and $\Tt,\Tt'\in\mbox{\rm U}(L,L,\CC)$ 
the M\"obius transformation $\Tt\cdot U$ is well-defined, is again in 
$\partial_l\DM_L^\KM $ and that
$(\Tt\Tt')\cdot U=\Tt\cdot(\Tt'\cdot U)$. Let $\Tt$ be given in terms of
$A,B,C,D$ as in Lemma~\ref{lem-Cayley}. Then this lemma implies that
$(CU+D)=D(\one+D^{-1}CU)$ is invertible so that the M\"obius transform is
well-defined. Let us first show that 
$(\Tt\cdot U)^*(\Tt\cdot U)\leq \one$. For this purpose, one can appeal to the
identity 
\begin{equation}
\label{eq-crux}
(C\,U\,+\,D)^*\,(C\,U\,+\,D)\;-\;
(A\,U\,+\,B)^*\,(A\,U\,+\,B)
\;=\;
\one\;-\;U^*\,U
\;,
\end{equation}
following directly from the identities in Lemma~\ref{lem-Cayley}. Indeed,
multiplying \eqref{eq-crux} from the left by
$(CU+D)^*)^{-1}$ and the right by $(CU+D)^{-1}$ and using 
$\one-U^*U\geq\nul$ shows $(\Tt\cdot U)^*(\Tt\cdot U)\leq \one$.
Next let us show that the invertible $(CU+D)$ maps ker$(\one-U^*U)$ to
ker$(\one-(\Tt\cdot U)^*(\Tt\cdot U))$ and
ker$(\one-U^*U)^\perp$ to
ker$(\one-(\Tt\cdot U)^*(\Tt\cdot U))^\perp$. If $v\in\,$ker$(\one-U^*U)$,
then \eqref{eq-crux} implies
$$
\|\,(C\,U\,+\,D)\,v\,\|
\;=\;
\|\,(A\,U\,+\,B)\,v\,\|
\;=\;
 \|\,\Tt\cdot U\,(C\,U\,+\,D)\,v\,\|
\;,
$$
so that $(CU+D)v\in\,$ker$(\one-(\Tt\cdot U)^*(\Tt\cdot U))$ because 
$\one-(\Tt\cdot U)^*(\Tt\cdot U)\geq 0$. Similarly for 
$v\in\,$ker$(\one-U^*U)^\perp$ one has 
$\|(CU+D)v\|>\|\Tt\cdot U(CU+D)v\|$ implying that 
$v\notin\,$ker$(\one-(\Tt\cdot U)^*(\Tt\cdot U))$.

The argument up to now shows that $\Tt\cdot U\in \partial_l\DM_L^\CC$. 
A short algebraic calculation also 
shows that $(\Tt\Tt')\cdot Z=\Tt\cdot(\Tt'\cdot Z)$. It
remains to show that the symmetries are conserved in the cases
$\KM=\RR,\HM$. For $\Tt\in U(L,L,\RR)$ one has $C=\overline{B}$ and
$D=\overline{A}$ by Lemma~\ref{lem-Cayley}, so that
$$
\Tt\cdot U
-
(\Tt\cdot U)^t
\;=\;
((\overline{B}U+\overline{A})^{-1})^t
\left[
(\overline{B}U+\overline{A})^t
(AU+B)-(AU+B)^t
(\overline{B}U+\overline{A})
\right]
(\overline{B}U+\overline{A})^{-1}
\;.
$$
For symmetric $U$ one checks that the term in the bracket vanishes due to the
identities in Lemma~\ref{lem-Cayley}, implying that $\Tt\cdot U$ is again
symmetric. Similarly one proceeds in the case $\KM=\HM$.

Now let us come to the last point of the proposition.
Given $U\in\partial_L\DM^\KM_L$, 
let $[\Phi]_\sim\in\LM_L^\KM$ 
be such that $\Pi([\Phi]_\sim)=U$ (by the
construction in the proof of Theorem~\ref{theo-realLag}). For 
$\Tt\in\,\mbox{\rm SP}(2L,\KM)$, one then has 
$[\Tt\Phi]_\sim\in\LM_L^\KM$. Theorem~\ref{theo-realLag} implies that both  
$[\Cc\Phi]_\sim$ and $[\Cc\Tt\Phi]_\sim=[\Cc\Tt\Cc^*\Cc\Phi]_\sim$ 
are in $\GM_L^\inv$. From \eqref{eq-implement} now follows that
$\Cc\Tt\Cc^*\cdot\pi([\Cc\Phi]_\sim)=\pi([\Cc\Tt\Phi]_\sim)$. 
\hfill $\Box$

\vspace{.2cm}

It is interesting to note (and relevant for Section~\ref{sec-Green})
that the M\"obius
transformation sends $\UM_L^\KM$ to $\UM_L^\KM$ for some
matrices in $\mbox{GL}(2L,\CC)$ which are not in 
$\mbox{SP}(2L,\KM)$. In particular, for $\delta>0$ one
even has:
\begin{equation}
\label{eq-energymoeb}
\left(
\begin{array}{cc} {\bf 1} & \imath\,\delta\,{\bf 1} \\  
{\bf 0} & {\bf 1} \end{array}
\right)
\cdot Z
\;=\;
Z+\imath\,\delta\,{\bf 1}
\;\in\;\UM_L^\KM\;,
\qquad
\mbox{ for }\;Z\in\UM^\KM_L\,\cup\;\partial \UM_L^\KM
\;.
\end{equation}

\vspace{.2cm}

The following formula shows that the M\"obius transformation appears naturally
in the calculation of a volume distortion by an invertible matrix, namely the
so-called Radon-Nykodym cocycle. It will be used for the calculation of the
sum of Lyapunov exponents in Section~\ref{sec-DOSformula}.

\begin{lemma}
\label{lem-RNcocycle}
Suppose that $[\Phi]_{\sim}\in\GM_L^\inv$ and
$[\Tt\Phi]_{\sim}\in\GM_L^\inv$ where $\Tt\in\,\mbox{\rm GL}(2L,\CC)$. With
notations for $\Tt$ as in {\rm \eqref{eq-moebius}} one has
$$
\frac{\det\bigl(\,(\Tt\Phi)^*(\Tt\Phi)\,\bigr)}{\det\bigl(\,\Phi^*\Phi\,\bigr)}
\;=\;
\frac{\det\bigl((\Tt\cdot\pi([\Phi]_\sim))^*(\Tt\cdot\pi([\Phi]_\sim))
+\one\bigr)}{\det\bigl((\pi([\Phi]_\sim))^*(\pi([\Phi]_\sim))
+\one\bigr)}
\;
\Bigl|\,\det\bigl(\,
C\,\pi([\Phi]_\sim)+D\,\bigr)\,\Bigr|^2
\;.
$$
\end{lemma}

\noindent {\bf Proof.} 
Let $\Phi=\left(\begin{array}{c} a \\  b \end{array}\right)$. As $b$ is
invertible by hypothesis, 
$$
\frac{\det\bigl(\,(\Tt\Phi)^*(\Tt\Phi)\,\bigr)}{\det\bigl(\,\Phi^*\Phi\,\bigr)}
\;=\;
\frac{\det\bigl(\,(Aab^{-1}+B)^*\,(Aab^{-1}+B)
+(Cab^{-1}+D)^*\,(Cab^{-1}+D)\,\bigr)}{
\det\bigl(\,(ab^{-1})^*\,(ab^{-1})+\one\,)}
\;.
$$
Furthermore, it was supposed that $Cab^{-1}+D$ is also
invertible, and this allows to conclude the proof because 
$\pi([\Phi]_\sim)=ab^{-1}$.
\hfill $\Box$

\subsection{Singular cylces}
\label{sec-singcyc}

Given $\xi\in\,$Sym$(L,\RR)\,\cap\,$Self$(L,\CC)$, let us set $\Psi_\xi=
\left(\begin{array}{c} \xi \\ \one \end{array}\right)$. 
The associated singular cycle (often called Maslov cycle, but it actually
already appears in Bott's work \cite{Bot}) is
$$
\LM_L^{\KM,\xi}
\;=\;
\left\{
[\Phi]_\sim\in\LM_L^\KM
\;\left|
\;
\Phi\;\CC^L\,\cap\,\Psi_\xi\,\CC^L
\;\neq\; 
\{\vec{0}\}
\right.\right\}
\;.
$$
It can be decomposed into a disjoint union of $\LM_L^{\KM,\xi,l}$,
$l=1,\ldots,L$, where 
$$
\LM_L^{\KM,\xi,l}
\;=\;
\left\{
[\Phi]_\sim\in\LM_L^\KM
\;\left|
\;
\dim\left(\Phi\;\CC^L\,\cap\,
\Psi_\xi\,\CC^L\right)
\;=\;l 
\right.\right\}
\;.
$$
It is possible to 
define singular cycles associated to Lagrangian planes which
are not of the form of
$\Psi_\xi$, but this will not be used here. 

\vspace{.2cm}

It is convenient to express the intersection conditions in terms of the
Wronskian associated to two
$2L\times L$ matrices $\Phi$
and $\Psi$ representing two Lagrangian planes 
$$
W(\Phi,\Psi)
\;=\;
\Phi^* {\cal J}\Psi
\;,
$$
namely one checks that $\Phi\,\CC^L\,\cap\,
\Psi\,\CC^L
\,\neq\,\{\vec{0}\}\;\Leftrightarrow\;
\det\bigl(\,W(\Phi,\Psi)\,\bigr)
\,=\,0\,$,
and, more precisely, 
\begin{equation}
\label{eq-interseccond2}
\dim
\bigl(\Phi\,\CC^L\,\cap\,
\Psi\,\CC^L\bigr)
\;=\;
\dim\bigl(
\mbox{ker}\,W(\Phi,\Psi)\,\bigr)
\;.
\end{equation}
Note that even though $W(\Phi,\Psi)$ 
does depend on the choice of  basis in the two Lagrangian planes, the
dimension of its kernel is independent of this choice. Furthermore,
$W(\Phi,\Psi)=\nul$ if and only if  $[\Phi]_\sim =[\Psi]_\sim$.

\vspace{.2cm}

Arnold showed in \cite{Arn} that 
$\LM_L^{\RR,\xi}$ is two-sided by exhibiting  a non-vanishing
transversal vector field on $\LM_L^{\RR,\xi}$. This allows to define a
weighted intersection number for paths in a generic position, namely for paths
having only intersections with the highest stratum $\LM_L^{\RR,\xi,1}$.
Bott proved a similar result for  $\LM_L^{\CC,\xi}$ already earlier \cite{Bot}.
Using the following proposition, 
it will be straightforward in the next section to define intersection numbers
for paths 
which are not necessarily in a generic position.

\begin{proposi}
\label{prop-sing}
$\Pi(\LM_L^{\KM,\xi,l})=\partial_L^{\xi,l}\DM_L^\KM$ 
where $\partial_L^{\xi,l}\DM_L^\KM$ is the following subset of the
maximal boundary $\partial_L\DM_L^\KM$:
$$
\partial_L^{\xi,l}\DM_L^\KM
\;=\;
\left\{\,
U\in
\partial_L\DM_L^\KM
\;\left|
\;
\mbox{\rm rank}\left(
\Pi([\Psi_\xi]_\sim)
-U
\right)
\;=\;L-l 
\right.
\right\}
\;.
$$
Setting $\partial_L^{\xi}\DM_L^\KM
=\cup_{l=1,\ldots,L}\,\partial_L^{\xi,l}\DM_L^\KM$,
one has $ \Pi(\LM_L^{\KM,\xi})=\partial_L^{\xi}\DM_L^\KM$. 
One has $\one\notin
\partial_L^{\xi}\DM_L^\KM $ 
for any $\xi \in\,${\rm Sym}$(L,\RR)\,\cap\,${\rm Self}$(L,\RR)$.
\end{proposi}

\noindent {\bf Proof.} 
Given $U$, let $\Phi=\left(\begin{array}{c} a \\  b \end{array}\right)$
where $a=\frac{1}{2}(U+\one)$ and $b=\frac{\imath}{2}(U-\one)$ as in the proof
of Theorem~\ref{theo-realLag}.
Hence $\Pi([\Phi]_\sim)=U$. Then one
verifies 
$$
W(\Psi_\xi,\Phi)
\;= \;
a\,-\,\xi\,b
\;=\;
\frac{1}{2}\;
\bigl(U\,+\,\one\,-\,\imath\,\xi\,(U\,-\,\one)\bigr)
\; = \;
\frac{1}{2}\,(\one-\imath\,\xi)\,
\bigl(
U\,-\,\Pi([\Psi_\xi]_\sim)
\bigr)
\;.
$$
As $\one-\imath\,\xi$ is invertible, the result follows directly from
\eqref{eq-interseccond2}. No more care is needed in the real case because the
dimension of the kernel of the Wronskian is independent of a basis change from
the above $\Phi$ to the real one in the proof of
Theorem~\ref{theo-realLag}. 
\hfill $\Box$

\vspace{.2cm}

We will only use the singular cycles associated to
$\xi=-\cot(\frac{\varphi}{2})\, \one$ where $\varphi\in (0,2\pi)$, and only
need to use $\KM=\CC$ as the supplementary symmetries are irrelevant for the
definition and calculation of the intersection number
(Bott or Maslov index) in the next sections. Let us
denote these singular cycles, subsets of $\LM^\CC_L$, 
by $\LM_L^{\varphi}= \cup_{l=1,\ldots,L}
\LM_L^{\varphi,l}$. The image under $\Pi$ will then be written as
$\partial_L^{\varphi}\DM_L^\CC= \cup_{l=1,\ldots,L}
\partial_L^{\varphi,l}\DM_L^\CC$.
Because  $\Pi(\Psi_\xi)=e^{\imath\varphi}\one$, it
follows from Proposition~\ref{prop-sing} that
\begin{equation}
\label{eq-specialsing}
\Pi(\LM_L^{\varphi,l})
\;=\;
\partial_L^{\varphi,l}\DM_L^\CC
\;=\;
\left\{
\,U\in
\,\mbox{U}(L)
\;\left|
\;
e^{\imath\varphi} 
\mbox{ eigenvalue of } U \mbox{ with multiplicity } l
\right.
\right\}
\;.
\end{equation}

\subsection{The intersection number (Bott index)}
\label{sec-Maslov}

Let $\Gamma=([\Phi^E]_\sim)_{E\in[E_0,E_1)}$ be a (continuous) closed
path in $\LM_L^\CC$ for which the number of intersections  $\{E\in
[E_0,E_1)\;|\;\Gamma(E)\in\LM_L^{\varphi}  \}$
is finite. 
At an intersection $\Gamma(E)\in\LM_L^{\varphi,l}$, 
let $\theta_1(E'),\ldots,\theta_l(E')$
be those eigenphases of the unitary $\Pi(\Gamma(E'))$ which are
all equal to $\varphi$ at
$E'=E$. Choose $\epsilon,\delta>0$ such that
$\theta_k(E')\in[\varphi-\delta,\varphi+\delta]$ for $k=1,\ldots,l$ and
$E'\in [E-\epsilon,E+\epsilon]$ and that there are no other eigenphases in 
$[\varphi-\delta,\varphi+\delta]$ for $E'\neq E$ and finally
$\theta_k(E')\neq\varphi$ for those parameters.
Let $n_-$ and $n_+$ be the number of those of the $l$ 
eigenphases less than $\varphi$
respectively before and after the intersection, and similarly let $p_-$ and
$p_+$ be the number of eigenphases larger than $\varphi$ before and after the
intersection. Then the signature of $\Gamma(E)$ is defined by
\begin{equation}
\label{eq-signature}
\mbox{\rm sgn}(\Gamma(E))
\;=\;
\frac{1}{2}\;(p_+-n_+-p_-+n_-)
\;=\;
l-n_+-p_-
\;.
\end{equation}
Note that $-l\leq\mbox{\rm sgn}(\Gamma(E))\leq l$ and that
$\mbox{\rm sgn}(\Gamma(E))$ is 
the effective number of eigenphases that have crossed
$\varphi$ in the counter-clock sense. Furthermore the signature is stable
under perturbations of the path
in the following sense: if an intersection by $\LM_L^{\varphi,l}$ is
resolved by a perturbation into a series of intersections by lower strata,
then the sum of their signatures is equal to $\mbox{\rm sgn}(\Gamma(E))$. 
Finally let us remark that, if the phases are differentiable and $\partial_E
\theta_k(E)\neq 0$ for $k=1,\ldots,l$, then $\mbox{\rm sgn}(\Gamma(E))$ is
equal to the sum of the $l$ signs $\mbox{\rm sgn}(\partial_E
\theta_k(E))$, $k=1,\ldots,l$. Now the intersection number or 
index of the path $\Gamma$ 
w.r.t. the singular cycle $\LM_L^{\varphi}$ is defined by 
\begin{equation}
\label{eq-intersec}
\mbox{\rm ind}(\Gamma,\LM_L^{\varphi})
\;=\;
\sum_{\Gamma(E)\,\in\,\LM_L^{\varphi}}
\;\mbox{\rm sgn}(\Gamma(E))
\;.
\end{equation}

Let us give a different expression for this index. If
$E\in[E_0,E_1]\mapsto \theta_l^E$ are continuous paths of the eigenphases of
$\Pi(\Gamma(E))$ with arbitrary choice of enumeration at level crossings, then
each of them leads to a winding number. A bit of thought shows that
\begin{equation}
\label{eq-intersec2}
\mbox{\rm ind}(\Gamma,\LM_L^{\varphi})
\;=\;
\sum_{l=1}^L
\;\mbox{Wind}
\Bigl(\;
E\in [E_0,E_1) \mapsto  \theta_l^E
\;\Bigr)
\;.
\end{equation}
In particular, the r.h.s. is independent of the choice of
enumeration at level crossings. 
Moreover, for the r.h.s. to make sense, one does not need to
impose that the number of intersections is finite, as is, of course, necessary
in order to define an intersection number. Similarly, the index of a closed 
path $\Gamma$ in the real or quaternion 
Lagrangian Grassmannian could be defined; however, this index 
coincides with $\mbox{\rm ind}(\Gamma,\LM_L^{\varphi})$ if $\Gamma$ is
considered as path in the complex Lagrangian Grassmannian.

\vspace{.2cm}

In the literature, $\mbox{\rm ind}(\Gamma,\LM_L^{\varphi})$ is often referred
to as the Maslov index, at least in the case of a path in $\LM^{\RR}_L$. The
same object already appears in the work of Bott \cite{Bot} though, and it
seems more appropriate to associate his name to it. The above definition using
\eqref{eq-signature} appears to be considerably more simple, and the author
does not know whether it was already used elsewhere. 

\subsection{Arnold's cocycle}
\label{sec-Arnold}

Arnold has shown \cite{Arn} that $H^1(\LM_L^\RR,\ZZ)\cong \ZZ$. This and 
$H^1(\LM^\KM_L,\ZZ)\cong \ZZ$ follows from Theorem~\ref{theo-realLag}. 
The generator $\omega$ of the de Rahm groups 
can be chosen as follows. A continuous closed path
$\Gamma=([\Phi^E]_\sim)_{E\in[E_0,E_1)}$ in $\LM_L^\CC$
gives rise to a continuous closed path 
$E\in [E_0,E_1) \mapsto \det(\Pi([\Phi^E]_\sim))$ in $S^1$. Its winding number
defines the pairing of (the de Rahm class of) $\omega$ with the (homotopy
class of the) path $\Gamma$:
$$
\langle\, \omega\,|\,\Gamma\,\rangle
\;=\;
\mbox{Wind}
\Bigl(\;
E\in [E_0,E_1) \mapsto \det(\Pi([\Phi^E]_\sim))
\;\Bigr)
\;.
$$
Any continuous path in $\LM_L^\CC$ can
be approximated by a differentiable one, hence we suppose from now on that 
$\Gamma$ is differentiable. Then one can calculate the pairing by
\begin{equation}
\label{eq-windint}
\langle\, \omega\,|\,\Gamma\,\rangle
\; = \;
\int_{E_0}^{E_1} \frac{dE}{2\pi}\;\,
\Im m\;\,\partial_E\;\log(\det(\Pi([\Phi^E]_\sim)))
\;.
\end{equation}
The next theorem is
Arnold's main result concerning this cocycle. 

\begin{theo}
\label{theo-Arnold} {\rm \cite{Arn}}
Provided a closed differentiable path 
$\Gamma$ has only finitely many intersections with 
the singular cycle $\LM_L^{\varphi}$, 
$$
\mbox{\rm ind}(\Gamma,\LM_L^{\varphi})
\;=\;
 \langle\, \omega\,|\,\Gamma\,\rangle
\;.
$$
\end{theo}

\noindent {\bf Proof.} Set $U(E)=\Pi(\Gamma(E))$.
As already used in the proof of Theorem~\ref{theo-realLag}, the
diagonalization $U(E)=M(E)^*D(E)M(E)$ can be done with
a differentiable unitary matrix $M(E)$ and a differentiable
diagonal matrix $D(E)=\mbox{diag}
(e^{\imath\theta^E_{1}},\ldots,e^{\imath\theta^E_{L}})$. 
As $M^E(\partial_EM^E)^* =-(\partial_E M^E)(M^E)^*$,
one has
$$
\Im m\;\,\partial_E\;\log(\det(U(E)))
\;=\;
\frac{1}{\imath}\;
\Tr\left(
(D(E))^*\partial_E D(E)\right)
\;=\;
\sum_{l=1}^L\;\partial_E\,\theta^E_{l}
\;.
$$
Integrating w.r.t. $E$ as in \eqref{eq-windint}
hence shows that 
$\langle\, \omega\,|\,\Gamma\,\rangle$ is equal to the sum of the winding
numbers of the eigenphases and this is equal to the index by
\eqref{eq-intersec2}.  
\hfill $\Box$

\vspace{.2cm}

\noindent {\bf Remark} It follows from the Gohberg-Krein index theorem that
the intersection number is also equal to the Fredholm index of an associated
Toeplitz operator \cite{BS}.

\vspace{.2cm}

If $\Tt\in\,$SP$(2L,\CC)$, 
then one can define another closed path in
$\LM_L^\CC$  by
$\Tt\Gamma=([\Tt\Phi^E]_\sim)_{E\in[E_0,E_1)}$. 
Because SP$(2L,\CC)$ is arc-wise
connected, it follows from the homotopy invariance of the pairing that
$\langle\, \omega\,|\,\Tt\Gamma\,\rangle=
\langle\, \omega\,|\,\Gamma\,\rangle$. The
following proposition allows $\Tt$ to vary and also
analyzes intermediate values of the
integral in \eqref{eq-windint}, denoted by:
$$
\int^E_{\Gamma}\,\omega
\;=\;
\int_{E_0}^{E} \frac{de}{2\pi}\;\,
\Im m\;\,\partial_e\;\log(\det(\Pi([\Phi^e]_\sim)))
\;,
\qquad
E\in[E_0,E_1)\;.
$$
%

\begin{proposi}
\label{prop-basischange}
Let $\Gamma=([\Phi^E]_\sim)_{E\in[E_0,E_1)}$ be a closed differentiable path
in $\LM_L^\CC$ and $(\Tt^E)_{E\in[E_0,E_1)}$ be a differentiable path in 
{\rm SP}$(2L,\CC)$ such that $\Gamma'=([\Tt^E\Phi]_\sim)_{E\in[E_0,E_1)}$ is a
closed path in $\LM_L^\CC$ for any given $[\Phi]_\sim\in \LM_L^\CC$.
Furthermore let us introduce the closed path
$\Gamma''=([\Tt^E\Phi^E]_\sim)_{E\in[E_0,E_1)}$. Then
$$
\langle\, \omega\,|\,\Gamma''\,\rangle
\;=\;
\langle\, \omega\,|\,\Gamma\,\rangle
\;+\;
\langle\, \omega\,|\,\Gamma'\,\rangle
\;.
$$
Furthermore, with the notation
$$
\Cc\,\Tt^E\,\Cc^*
\;=\;
\left(
\begin{array}{cc} A^E & B^E \\  C^E & D^E \end{array}
\right)
\;,
$$
one has {\rm (}independent of $[\Phi]_\sim${\rm )}
\begin{equation}
\label{eq-symprot}
\langle\, \omega\,|\,\Gamma'\,\rangle
\;=\;
\int_{E_0}^{E_1}
\frac{dE}{2\,\pi}\;
\Im m\;\,\partial_E\;\log\bigl(\det(A^E(D^E)^{-1})\bigr)
\;,
\end{equation}
and, uniformly in $E$,
$$
\left| \;
\int^E_{\Gamma''}\,\omega\;-\;
\int^E_{\Gamma}\,\omega
\;-\;
\int_{E_0}^{E}
\frac{de}{2\,\pi}\;
\Im m\;\,\partial_e\;\log\bigl(\det(A^e(D^e)^{-1})\bigr)
\;\right|
\;\leq\;L\;.
$$
\end{proposi}

\noindent {\bf Proof.} Set $U^e=\Pi([\Phi^e]_\sim)$. 
Due to Theorem~\ref{theo-Usymdyn}, 
$$
\int^E_{\Gamma''}\,\omega
\;=\;
\int_{E_0}^{E} \frac{de}{2\pi}\;\,
\Im m\;\,\partial_e\,
\Bigl(\;
\log(\det(A^e\,U^e+B^e))
-
\log(\det(C^e\,U^e+
D^e\,))
\;
\Bigr)
\;.
$$
In the first contribution, let us use
$\det(A^e\,U^e+B^e)=\det(B^e\,(U^e)^*+A^e)\det(U^e)$, 
of which the second factor leads
to  $\int^E_{\Gamma}\,\omega$.  Hence
$$
\int^E_{\Gamma''}\,\omega
\;-\;
\int^E_{\Gamma}\,\omega
\;=\;
\int_{E_0}^{E} \frac{de}{2\,\pi}\;\,
\Im m\;\,\partial_e\,
\left(
\log(\det(B^e\,(U^e)^*+A^e))-
\log(\det(C^e\,U^e+
D^e\,))
\right)
\;.
$$
Now $A^e$ and $D^e$ are invertible due to Lemma~\ref{lem-Cayley}. Hence
\begin{eqnarray*}
\int^E_{\Gamma''}\omega\!\!
& - &
\!\!\int^E_{\Gamma}\omega
\,-\,
\int_{E_0}^{E} \frac{de}{2\,\pi}\;\,
\Im m\;\,\partial_e\,\log\bigl(\det(A^e(D^e)^{-1})\bigr)
\\
& = &
\int_{E_0}^{E} \frac{de}{2\,\pi}\;\partial_e\;
\Im m\;
\Bigl(\;
\Tr\left(\log(\one+(A^e)^{-1}B^e\,(U^e)^*)\right)
-
\Tr\left(\log(\one+(D^e)^{-1}C^e\,U^e)\right)
\;\Bigr)
\;.
\end{eqnarray*}
As $\|(A^e)^{-1}B^e\|<1$ and $\|(D^e)^{-1}C^e\|<1$
by Lemma~\ref{lem-Cayley}, only one branch of
the logarithm is needed. As $E\to E_1$ this term therefore vanishes implying
the result on the winding numbers because the third term on the l.h.s. then
gives $\langle\, \omega\,|\,\Gamma'\,\rangle$ as one sees repeating the above
arguments with $U^e$ replaced by the constant $\Pi([\Phi]_\sim)$.  Moreover,
one may carry out the integral on the r.h.s. of the last equation 
using the fundamental theorem, so that this r.h.s. is equal to
$$
\frac{1}{2\,\pi}\;
\Bigl(\;
\Tr\left(\log(\one+(A^E)^{-1}B^E\,(U^E)^*)\right)
-
\Tr\left(\log(\one+(D^E)^{-1}C^E\,U^E)\right)
\;\Bigr)
\;.
$$
The bound follows now from the spectral mapping theorem for the logarithm
function. 
\hfill $\Box$

\vspace{.2cm}

In order to calculate the integral in \eqref{eq-windint}, 
one can also appeal to the following formula.

\begin{lemma}
\label{lem-derivative}
If $E\mapsto \Phi^E$ is differentiable and $[\Phi^E]_\sim\in\LM_L^\CC$, then
\begin{equation}
\label{eq-derivative}
\Im m\;\,\partial_E\;\log(\det(\Pi([\Phi^E]_\sim)))
\;=\;
2\;
\Tr\Bigl(\;
\bigl((\Phi^E)^* \Phi^E\bigr)^{-1}\; 
(\Phi^E)^*\,\Jj\,(\partial_E\,\Phi^E)
\Bigr)
\;.
\end{equation}
\end{lemma}

\noindent {\bf Proof.} 
As in the proof of Theorem~\ref{theo-Arnold},
let us begin with the identity
\begin{equation}
\label{eq-help1}
\imath\;\partial_E\;\log(\det(\Pi([\Phi^E]_\sim)))
\; = \;
\Tr\Bigl(\;
(\Pi([\Phi^E]_\sim))^*\;
(\partial_E\,\Pi([\Phi^E]_\sim))\;
\Bigr)
\;.
\end{equation}
Introducing the invertible $L\times L$ matrices
$$
\phi^E_\pm
\;=\;
(\,\one\;\pm\!\imath\one\,)\;\Phi^E
\;,
$$
one has $\Pi([\Phi^E]_\sim)=\phi^E_-\,(\phi^E_+)^{-1}$. Furthermore the
following identities can be checked using the fact that $\Phi^E$ is Lagrangian:
\begin{equation}
\label{eq-help2}
(\phi^E_+)^*\,\phi^E_+
\;=\;
(\phi^E_-)^*\,\phi^E_-
\;=\;
(\Phi^E)^*\,\Phi^E
\;,
\qquad
(\phi^E_\pm)^*\partial_E\phi^E_\pm
\;=\;
(\Phi^E)^*\partial_E\Phi^E
\,\pm\,\imath\,
(\Phi^E)^*\Jj\partial_E\Phi^E
\,.
\end{equation}
Corresponding to the two terms in
$$
\partial_E\,\Pi([\Phi^E]_\sim)
\;=\;
(\partial_E\phi^E_-)\,(\phi^E_+)^{-1}
\,-\,
\phi^E_-\,(\phi^E_+)^{-1}(\partial_E\phi^E_+)\,(\phi^E_+)^{-1}
\;,
$$
there are two contributions $C_1$ and $C_2$ in \eqref{eq-help1}. For the first
one, it follows from cyclicity
$$
C_1
\;=\;
\Tr\Bigl(\;
\bigl((\phi^E_+)^* \phi^E_+\bigr)^{-1}\; 
(\phi^E_-)^*\,\partial_E\,\phi^E_-\,
\Bigr)
\;,
$$
while for the second
$$
C_2
\; = \;
-\,
\Tr\Bigl(\;
(\phi^E_+)^{-1}\,\partial_E\,\phi^E_+\,
\Bigr)
\;=\;
-\,
\Tr\Bigl(\;
\bigl((\phi^E_+)^* \phi^E_+\bigr)^{-1}\; 
(\phi^E_+)^*\,\partial_E\,\phi^E_+\,
\Bigr)
\;.
$$
Combining them and appealing to \eqref{eq-help2} concludes the proof.
\hfill $\Box$

\vspace{.2cm}

Next let us calculate the pairing for two examples. 
The first one was already given
by Arnold \cite{Arn}. For $\eta\in [0,\pi]$, introduce the
symplectic matrices
$$
R_\eta
\;=\;
\left(
\begin{array}{cc} \cos(\eta)\,\one & \sin(\eta)\,\one 
\\ -\,\sin(\eta)\,\one & \cos(\eta)\,\one \end{array}
\right)
\;,
\qquad
\Cc\, R_\eta\,\Cc^*
\;=\;
\left(
\begin{array}{cc} e^{\imath\eta}\,\one & \nul 
\\ \nul & e^{-\imath\eta}\, \one \end{array}
\right)
\;,
$$
as well as the closed path $\Gamma=
(R_\eta[\Phi]_\sim)_{\eta\in[0,\pi)}$ for an arbitrary 
$[\Phi]_\sim\in\LM^\CC_L$. As $\det(\Pi([R_\eta\Phi]_\sim))=e^{2\imath L\eta}
 \det(\Pi([\Phi]_\sim))$, one deduces 
$\langle\, \omega\,|\,\Gamma\,\rangle=L$. The second example concerns transfer
matrices. 

\begin{lemma}
\label{lem-transferwind}
Let $\Tt^E$ be a matrix built as in {\rm \eqref{eq-transfer}} from a 
selfadjoint matrix $V$ and posi\-tive matrix $T$.
For an arbitrary $[\Phi]_\sim\in\LM^\CC_L$ with $(\one\,\nul)\,\Phi$ invertible
and $\Mm\in\,${\rm SP}$(2L,\CC)$, consider the path 
$\Gamma=(\Mm\Tt^E[\Phi]_\sim)_{E\in\overline{\RR}}$ 
where $\overline{\RR}=\RR\cup\{\infty\}$ is the
one-point compactification. Then $\Gamma$ is closed and
$\langle\, \omega\,|\,\Gamma\,\rangle=L$. Moreover, 
$E\in\RR\mapsto\int^E_\Gamma\omega$ is strictly monotonously increasing.
\end{lemma}

\noindent {\bf Proof.} First suppose that $\Mm=\one$. Then 
note that $\Pi([\Tt^E\Phi]_\sim)=
\one+\Oo(E^{-1})$, hence the path is closed.
Next one calculates
$$
\Cc\,\Tt^E\,\Cc^*
\;=\;
\frac{1}{2}
\left(
\begin{array}{cc}
(E\,{\bf 1}\,-\,V)\,T^{-1} 
-\imath(T+T^{-1})
\;\; &  (E\,{\bf 1}\,-\,V)\,T^{-1} 
+\imath(T-T^{-1})
\\ (E\,{\bf 1}\,-\,V)\,T^{-1} 
-\imath(T-T^{-1})
\;\; & (E\,{\bf 1}\,-\,V)\,T^{-1} 
+\imath(T+T^{-1})
\end{array}
\right)
\;.
$$
Let $A^E$ and $D^E$ denote the upper
left and lower right
entry. Then $\Im m\;
\partial_E\,\log(\det(A^E))$ is equal to
$$
\Tr\left(
\bigl(T^{-1}(E\,\one-V)^2T^{-1}+(T+T^{-1})^2+\imath(T^{-1}VT-TVT^{-1})\bigr)^{-1}
(\one+T^{-2})
\right)\;.
$$
As there is a similar expression for $\Im m\;
\partial_E\,\log(\det(D^E))$, this
shows that the integral in \eqref{eq-symprot} is finite. In order to
calculate the winding number, it is convenient to consider the homotopy 
$$
\Tt^E(\lambda)
\;=\;
\left(
\begin{array}{cc}
(E\,{\bf 1}\,-\,V(\lambda))\,T(\lambda)^{-1} & - T(\lambda) \\
T(\lambda)^{-1} & {\bf 0}
\end{array}
\right)
\;,
\qquad
0\leq \lambda\leq 1
\;.
$$
where $V(\lambda)=\lambda
V$ and $T(\lambda)=\lambda T+(1-\lambda)\one$ (the latter is always positive). 
Then the pairing of
$\Gamma(\lambda)=(\Tt^E(\lambda)[\Phi]_\sim)_{E\in\overline{\RR}}$
with $\omega$ is independent of $\lambda$. Hence it is sufficient to calculate
the pairing at $\lambda=0$, which is due to the above replaced into
\eqref{eq-symprot} given by
$$
\langle\, \omega\,|\,\Gamma(0)\,\rangle
\;=\;
\int^\infty_{-\infty}\frac{dE}{\pi}\;
\Tr\bigl((E^2+4)^{-1}\,2\;\one\bigr)
\;=\;L
\;,
$$
completing the proof in the case $\Mm=\one$. If $\Mm\neq\one$, the winding
number is the same because SP$(2L,\CC)$ is arc-wise
connected, as already pointed out above. The monotonicity 
can be checked using the r.h.s. of \eqref{eq-derivative} for
$\Phi^E=\Mm\Tt^E\Phi$. In fact, $(\Phi^E)^*\Phi^E$ is strictly
positive, and 
$$
(\Phi^E)^*\Jj\partial_E\Phi^E
\;=\;
\Phi^*
\,
\left(
\begin{array}{cc}
T^{-2} & \nul \\
\nul & \nul
\end{array}
\right)
\,\Phi\;,
$$
which is strictly positive by hypothesis. As
the trace of a product of two positive
operators is still positive, this concludes the proof.
\hfill $\Box$ 

\section{Jacobi matrices with matrix entries}
\label{sec-JMME}

Given integers $L,N\in \NN$, 
let $(T_n)_{n=2,\ldots,N}$ and $(V_n)_{n=1,\ldots,N}$ 
be sequences of respectively positive
and selfadjoint $L\times L$ matrices
with complex entries. Furthermore let the left and right boundary conditions 
$\zeta$ and $\xi$ be also self-adjoint $L\times L$ matrices. 
In the real and quaternion case, one chooses
$\zeta$ and $\xi$  symmetric and self-dual. Then the associated Jacobi
matrix with matrix entries $H^N_\xi$ is by definition the symmetric 
operator acting on states $\phi=(\phi_n)_{n=1,\ldots,N}\in
\ell^2(1,\ldots,N)\otimes \CC^L$ by
\begin{equation}
\label{eq-jacobi}
(H^N_\xi\phi)_n
\;=\;
T_{n+1}\phi_{n+1}\,+\,V_n\phi_n
\,+\,T_{n}\phi_{n-1}
\;,
\qquad
n=1,\ldots,N\;,
\end{equation}
where $T_1=T_{N+1}={\bf 1}$,
together with the boundary conditions 
\begin{equation}
\label{eq-boundary}
\phi_0\;=\;\zeta\,\phi_1\;,
\qquad
\phi_{N+1}\;=\;\xi\,\phi_N\;.
\end{equation}
One can also rewrite $H^N_\xi$ defined in \eqref{eq-jacobi} and
\eqref{eq-boundary} more explicitly 
as a block diagonal matrix; this gives \eqref{eq-matrix} albeit with $V_1$ and
$V_N$ replaced by $V_1-\zeta$ and $V_N-\xi$.
The dependence on $\zeta$ is not specified, but it could potentially be used
for averaging purposes.
If $\zeta=\xi={\bf 0}$ one speaks of Dirichlet boundary conditions.

\subsection{Transfer matrices and dynamics of Lagrangian planes}
\label{sec-transfer}

As for a one-dimensional Jacobi matrix, 
it is useful to rewrite the Schr\"odinger
equation 
\begin{equation}
\label{eq-Schroedinger}
H^N_\xi\phi
\;=\;
E\,\phi\;,
\end{equation}
for a complex energy $E$
in terms of the $2L\times 2L$ transfer matrices $\Tt_n^E$ defined in
\eqref{eq-transfer}. For a real energy $E\in\RR$,
each transfer matrix is in the symplectic group SP$(2L,\CC)$. 
If $H^N$ is, moreover, real or self-dual, then the transfer matrices are 
in the subgroups   SP$(2L,\RR)$ and SP$(2L,\HM)$ respectively. The 
Schr\"odinger equation (\ref{eq-Schroedinger}) is satisfied if and only if
$$
\left(
\begin{array}{c}
T_{n+1}\phi_{n+1} \\
\phi_n
\end{array}
\right)
\;=\;
\Tt^E_n\,
\left(
\begin{array}{c}
T_{n}\phi_{n} \\
\phi_{n-1}
\end{array}
\right)
\;,
\qquad
n=1,\ldots,N
\;,
$$
and the boundary conditions (\ref{eq-boundary}) hold, namely
\begin{equation}
\label{eq-boundary2}
\left(
\begin{array}{c}
T_1\phi_{1} \\
\phi_0
\end{array}
\right)
\;\in\;
\Phi_\zeta
\,\CC^L
\;,
\qquad
\left(
\begin{array}{c}
T_{N+1}\phi_{N+1} \\
\phi_N
\end{array}
\right)
\;\in\;
\Psi_\xi
\,\CC^L
\;,
\end{equation}
where we introduced for later convenience the notations
$$
\Phi_\zeta
\;=\;
\left(
\begin{array}{c}
{\bf 1} \\
\zeta
\end{array}
\right)\;,
\qquad
\Psi_\xi
\;=\;
\left(
\begin{array}{c}
\xi \\
{\bf 1}
\end{array}
\right)
$$
Both of the two $L$-dimensional subspaces  of $\CC^{2L}$ 
appearing in the conditions \eqref{eq-boundary2} are
Lagrangian. One way to search for eigenvalues 
is to consider the whole 
subspace in the left equation 
of \eqref{eq-boundary2}, then to follow its evolution under application of the
transfer matrices, and finally to 
check whether at $N$ the resulting subspace has a non-trivial intersection
with the subspace on the r.h.s. of \eqref{eq-boundary2}.
For perturbation theory in Section~\ref{sec-covariant}, 
it is useful to incorporate a symplectic
basis change $\Mm\in\,$SP$(2L,\CC)$ which can be conveniently chosen later
on. In the one-dimensional situation this corresponds to pass to modified
Pr\"ufer variables. If $H^N$ is real or self-dual, 
one chooses $\Mm\in\,$SP$(2L,\RR)$ or $\Mm\in\,$SP$(2L,\HM)$.
As above, half-dimensional subspaces will be described by
$2L\times L$ matrices $\Phi^E_n$ of rank $L$ 
composed of column vectors spanning it. Then
their dynamics under application of the $\Mm$-transformed
transfer matrices is
\begin{equation}
\label{eq-Lagdyn}
\Phi_n^E
\;=\;
\Mm\,\Tt_n^E\,\Mm^{-1}\,\Phi_{n-1}^E
\;,
\qquad
\Phi_0^E
\;=\;
\Mm\,\Phi_\zeta
\;.
\end{equation}
If $E\in\RR$, these planes are Lagrangian.
As the boundary condition on the left boundary 
is satisfied automatically (it is chosen as the initial condition), 
the second condition in \eqref{eq-boundary2}
multiplied by $\Mm$
together with the Wronski test \eqref{eq-interseccond2} leads to
\begin{equation}
\label{eq-eigencond}
\mbox{multiplicity
of $E$ as eigenvalue of $H^N_\xi$}
\;=\;
L\,-\,\mbox{rank}
\left(\,(\Mm\,\Psi_\xi)^*
\;\Jj\;\Phi^E_N
\right)
\;.
\end{equation}
This implies also

\begin{proposi}
\label{prop-intersec} 
Let $\Gamma=([\Phi^E_N]_\sim)_{E\in\overline{\RR}}$. For every
$\xi\in\mbox{\rm Sym}(L,\RR)\,\cap\,\mbox{\rm Self}(L,\CC)$, the set 
$\{E\in\overline{\RR}\;|\;\Gamma(E)\in\LM_L^{\CC,\xi}  \}$
of intersections with the singular cycle $\LM_L^{\CC,\xi}$ is finite.
\end{proposi}

\vspace{.2cm}

The dynamics \eqref{eq-Lagdyn} 
is more easily controlled under the stereographic projection. 
Let us first consider the case  $\Im m(E)>0$. In this situation 
the stereographic projection $\pi$
of \eqref{eq-Lagdyn} gives a dynamics in the upper half-plane $\UM_L^\CC$, or
$\UM_L^\RR$ if $H^N$ is real and $\UM_L^\HM$ if $H^N$ is self-dual. 
In fact,  $Z_1^E=\pi(\Phi^E_1)=\Mm\cdot(E\,{\bf 1}-V_1-\zeta)$ is in
$\UM^\CC_L$ and, moreover, the transfer matrices factor as
follows: 
$$
\Tt_n^E
\;=\;
\left(
\begin{array}{cc}
{\bf 1}  &  \imath\,\Im m(E)\,{\bf 1} \\
{\bf 0} & {\bf 1}
\end{array}
\right)
\;\left(
\begin{array}{cc}
(\Re e(E)\,{\bf 1}\,-\,V_n)\,T_n^{-1} & - T_n \\
T_n^{-1} & {\bf 0}
\end{array}
\right)
\;.
$$
The matrix on the right is in the
symplectic group SP$(2L,\CC)$ acting on $\UM_L^\CC$, the one on the left also
sends $\UM_L^\CC$ to $\UM_L^\CC$ because of
\eqref{eq-energymoeb}. The same applies for real and self-dual $H^N$.
Hence the following M\"obius action is well-defined:
\begin{equation}
\label{eq-prueferdyn}
Z_n^E
\;=\;
\Mm\,\Tt_n^E\,\Mm^{-1}\cdot Z_{n-1}^E
\;,
\qquad
Z_1^E\;=\;\Mm\cdot(E\,{\bf 1}\,-\,V_1\,-\,\zeta)\;,
\end{equation}
In the case $\Mm=\one$, this is a matricial Ricatti equation 
$Z_n^E=E-V_n+T_n (Z_{n-1}^E)^{-1}T_n$.
Comparing with \eqref{eq-Lagdyn},
$$
Z_n^E\;=\;\pi([\Phi^E_n]_\sim)\;,
\qquad
n=1,\ldots,N\;.
$$
In particular, $[\Phi^E_n]_\sim$ is in the domain $\GM^\inv_L$ of $\pi$
whenever $\Im m(E)>0$.
If the boundary condition $\zeta$ is invertible, one may also set
$Z_0^E=\Mm\cdot\zeta^{-1}$ and then $Z^E_1=\Mm\Tt_1^E\Mm^{-1}
\cdot Z_0^E$, even though $Z_0^E$ is not in $\UM_L^\CC$. Furthermore, 
the map $E\in\UM_1^\CC\mapsto Z^E_N\in\UM_L^\CC$
is analytic (Herglotz). 
The map has
poles on the real axis as can be
read off the Dean-Martin identity \eqref{eq-DM} proven below, 
but will not be used in the sequel. 


\vspace{.2cm}

As real energies are not always permitted, 
the $Z^E_n$ are not convenient for the
calculation of the eigenvalues. For any $E$ with $\Im m (E)\geq 0$, 
let us rather use
$$
U^E_n
\;=\;
\Pi([\Phi^E_n]_\sim)
\;,
\qquad
n=0,\ldots,N
\;.
$$
For real $E$, this is well-defined and
$U^E_n$ is a unitary because of Theorem~\ref{theo-realLag}. This unitary
is symmetric or self-dual if $H^N$ is real or self-dual.
Iterating \eqref{eq-Lagdyn} and recalling the definition of the stereographic
projection shows that $U^E_n$ is actually of the explicit form given in
\eqref{eq-Uintro} if one chooses $\Mm=\one$. 
Hence this proves Theorem~\ref{theo-osci}(i) and part of (ii). 
For $\Im m(E)> 0$, the above arguments imply $Z^E_n$ is well-defined and
hence also $U^E_n=\Cc\cdot Z^E_n$. 
One, moreover, concludes that
$U^E_n=\Cc\cdot Z^E_n$ is in the generalized unit 
disc $\DM_L^\CC$ (for $n\neq 0$). 
The dynamics is given by
\begin{equation}
\label{eq-Usymdyn}
U_n^E
\;=\;
\Cc\,\Mm\,\Tt_n^E\,\Mm^{-1}\,\Cc^*\cdot U_{n-1}^E
\;,
\qquad
U_0^E\;=\Mm\cdot
({\bf 1}\,-\,\imath\,\zeta)\,({\bf 1}\,+\,\imath\,\zeta)^{-1}
\;.
\end{equation}
For $\Im m(E)> 0$, this is just the Cayley transform of 
\eqref{eq-prueferdyn}, while for $E\in\RR$, it is the dynamics of
Theorem~\ref{theo-Usymdyn}.  
The following lemma proves Theorem~\ref{theo-osci}(ii) and the first
part of (iii).

\begin{lemma}
\label{lem-analytic}
The map $E\mapsto U^E_N$ 
is analytic in a neighborhood of
$\overline{\UM_1^\CC}=\UM_1^\CC\,\cup\,\partial\UM_1^\CC$.  
At level crossings, the eigenvalues
and eigenvectors can be enumerated such that they are
analytic in a neighborhood of $\overline{\UM_1^\CC}$ as well.
\end{lemma}

\noindent {\bf Proof.} Analyticity of $U^E_N$
away from the real axis follows from the
analyticity of $Z^E_N$ because $U^E_n=\Cc\cdot Z^E_n$ and the inverse in the
M\"obius transformation is also well-defined, {\it cf.}
Section~\ref{sec-disc}. Moreover, 
the characteristic polynomial is a Weierstrass
polynomial that has a global 
Puiseux expansion which is analytic in the $L$th root of
$E$. Hence the eigenvalues and eigenvectors 
can be chosen (at level crossings) such that they are analytic in the
$L$th root of $E$ ({\it e.g.} \cite[Chapter II]{Kat}). 
It will follow from the arguments below that the Puiseux
expansion actually reduces to a power series expansion in $E$.

\vspace{.1cm}

Now we analyze in more detail the situation in a neighborhood of the real
axis. The plane $\Phi^E_N$ is a polynomial in $E$. Let us
use the notations
$\Phi^E_N=\left(\begin{array}{c} a^E \\  b^E \end{array}\right)$. It follows
from the argument in \eqref{eq-rank} that  $a^E +\imath\, b^E$ has maximal
rank for $E\in\RR$ so that $E\in\RR\mapsto \det(a^E +\imath\, b^E)$ has no
zero. Moreover, one has the large $E$ asymptotics
\begin{equation}
\label{eq-Phiasym}
\Phi^E_N
\;=\;
E^{N}\;\Mm\;
\left(
\begin{array}{c}
\prod_{n=1}^N T_n^{-1} \\
{\bf 0}
\end{array}
\right)
\;+\;
\Oo(E^{N-1})
\;.
\end{equation}
which implies
$$
\det(a^E +\imath\, b^E)
\;=\;
E^{NL}\;\det\left(
\left(
\begin{array}{c}
\one \\ \imath\,\one
\end{array}
\right)^t\,\Mm\,
\left(
\begin{array}{c}
\one \\ \nul
\end{array}
\right)
\right)\;
\prod_{n=1}^N\det(T_n)^{-1}
\;+\;
\Oo(E^{NL-1})
\;.
$$
Therefore $\inf_{E\in\RR}|\det(a^E +\imath\, b^E)|>0$ and the infimum is
realized at some finite $E$. A perturbative argument shows that also
$\inf_{E\in S_\delta}|\det(a^E +\imath\, b^E)|>0$
where $S_\delta=\{E\in\CC\,|\,|\Im m(E)|<\delta\}$ is a strip of some width
$\delta>0$. Calculating the inverse with the Laplace formula shows that
$(a^E +\imath\, b^E)^{-1}$ is also analytic in $S_\delta$. Thus also
$U^E_N=(a^E -\imath\, b^E)(a^E +\imath\, b^E)^{-1}$ is analytic in $S_\delta$
and thus analytic in a neighborhood of $\overline{\UM_1^\CC}$ due to the above.

\vspace{.1cm}

Finally one can appeal to degenerate perturbation theory
\cite[Theorem~II.1.10]{Kat} 
in order to deduce that the eigenvalues and eigenvectors
of the unitary matrix $U^E_N$ (hence $E$ real) are also 
analytic in a neighborhood of the real axis, that is are given by an analytic
Puiseux expansion. As this neighborhood has an open
intersection with the upper half-plane, the above Puiseux expansion
is therefore also analytic, namely only contains powers of $E$.
\hfill $\Box$

\vspace{.3cm}

It follows from \eqref{eq-Phiasym} that
\begin{equation}
\label{eq-Uasym}
U^E_N
\;=\;
\Cc\,\Mm\,\Cc^*\cdot\one\;+\;\Oo(E^{-1})
\;.
\end{equation}
Let $0\leq \theta^\Mm_l<2\pi$ be the eigenphases of the symmetric unitary
$\Cc\Mm\Cc^*\cdot\one$.
The eigenvalues of $U^E_N$, chosen to be real analytic in $E\in\RR$ as in
Lemma~\ref{lem-analytic}, are denoted by 
$e^{\imath\theta^E_{N,l}}$, $l=1,\ldots,L$. The eigenphases are chosen such
that $\theta^E_{N,l}\to \theta^\Mm_l$ for 
$E\to\,-\,\infty$.
In the case $\Mm=\one$, one hence has 
$\theta^E_{N,l}\to 0$ for $E\to\,-\,\infty$ 
as in Theorem~\ref{theo-osci}.

\vspace{.2cm}

Let us conclude this section by choosing particular right boundary
conditions, namely, for $\varphi\in (0,2\pi)$,
$$
\xi
\;=\;
-\,\cot(\frac{\varphi}{2})\;\Mm^{-1}\cdot\one\;
\qquad
\Longrightarrow
\qquad
\bigl[\Mm\Psi_\xi\bigr]_\sim
\;=\;
\bigl[\Psi_{-\cot(\frac{\varphi}{2})\,\one}\bigr]_\sim
\;.
$$
The
corresponding Hamiltonian will be denoted by $H^N_\varphi$. 
If $\Mm=\one$ and $\varphi=\pi$, these are Dirichlet boundary
conditions on the right boundary. 
Due to Proposition~\ref{prop-sing} and 
\eqref{eq-specialsing}, the eigenvalue condition \eqref{eq-eigencond} becomes
\begin{equation}
\label{eq-eigencond2}
\mbox{multiplicity
of $E$ as eigenvalue of $H^N_\varphi$}
\;=\;
\mbox{multiplicity
of $e^{\imath\varphi}$ as eigenvalue of $U^E_N$}
\;.
\end{equation}
\noindent Setting $\Mm=\one$ and 
$\varphi=\pi$, this proves Theorem~\ref{theo-osci}(iv).

\subsection{Monotonicity and transversality}

This section provides the proof of Theorem~\ref{theo-osci}(v) (just set
$\Mm=\one$ in the below). Of course, the second statement of 
Theorem~\ref{theo-osci}(v) follows immediately from the first one upon
evaluation in the eigenspaces of $U^E_N$. The following proposition also shows
that the curve $\Gamma=([\Phi^E_N]_\sim)_{E\in\overline{\RR}}$ 
is transversal to the singular cycle $\LM^{\varphi}_L$ 
and always crosses it from the negative to the positive
side. 

\begin{proposi}
\label{prop-derivbound}
For $E\in\RR$ and $N\geq 2$, one has
$$
\frac{1}{\imath}\,(U^E_N)^*\,\partial_E\,U^E_N
\;>\;0
\;.
$$
\end{proposi}

\noindent {\bf Proof.} 
As in the proof of Lemma~\ref{lem-derivative}, let us introduce
$\phi^E_\pm=(\,\one\;\pm\!\imath\one\,)\;\Phi^E_N$. These are
invertible $L\times L$ matrices and one has
$U^E_N=\phi_-^E(\phi_+^E)^{-1}=((\phi_-^E)^{-1})^*(\phi_+^E)^*$. Now
$$
(U^E_N)^*\,\partial_E\,U^E_N
\;=\;
((\phi_+^E)^{-1})^*
\Bigl[\,
(\phi_-^E)^*\partial_E\phi_-^E \,-\,
(\phi_+^E)^*\partial_E\phi_+^E
\,\Bigr]
(\phi_+^E)^{-1}
\;.
$$
Thus it is sufficient to verify positive definiteness of
$$
\frac{1}{\imath}\;
\Bigl[\,
(\phi_-^E)^*\partial_E\phi_-^E \,-\,
(\phi_+^E)^*\partial_E\phi_+^E
\,\Bigr]
\;=\;
2\;
(\Phi^E_N)^*\,\Jj\,\partial_E \Phi^E_N
\;.
$$
From the product rule follows that
$$
\partial_E \Phi^E_N
\;=\;
\sum_{n=1}^N
\;
\Mm\left(
\prod_{l=n+1}^N\,\Tt^E_l
\right)
\,
\left(\partial_E \Tt^E_n\right)
\;
\left(
\prod_{l=1}^{n-1}\,\Tt^E_l
\right)
\,\Phi_\zeta
\;.
$$
Using $\Mm^*\Jj\Mm=\Jj$, this implies that
\begin{equation}
\label{eq-help3}
(\Phi^E_N)^*\,\Jj\,\partial_E \Phi^E_N
\;=\;
\sum_{n=1}^N
\;\Phi_\zeta^*\,
\left(
\prod_{l=1}^{n-1}\,\Tt^E_l
\right)^*
\,
\bigl(\Tt^E_n\bigr)^*\,\Jj\,
\bigl(\partial_E \Tt^E_n\bigr)
\;
\left(
\prod_{l=1}^{n-1}\,\Tt^E_l
\right)
\,\Phi_\zeta
\;.
\end{equation}
As one checks that
$$
\bigl(\Tt^E_n\bigr)^*\,\Jj\,
\bigl(\partial_E \Tt^E_n\bigr)
\;=\;
\left(
\begin{array}{cc}
(T_nT_n^*)^{-1} & {\bf 0} \\
{\bf 0} & {\bf 0}
\end{array}
\right)
\;,
$$
each of the summands in \eqref{eq-help3} is positive semi-definite. 
In order to prove the strict inequality, it is sufficient that the first two
terms $n=1,2$ in \eqref{eq-help3} give a strictly positive contribution. Hence 
let us verify that
$$
\bigl(\Tt^E_2\bigr)^*\;
\left(
\begin{array}{cc}
(T_1T_1^*)^{-1} & {\bf 0} \\
{\bf 0} & {\bf 0}
\end{array}
\right)
\;\Tt^E_2
\;+\;
\left(
\begin{array}{cc}
(T_2T_2^*)^{-1} & {\bf 0} \\
{\bf 0} & {\bf 0}
\end{array}
\right)
\;>\;
0
\;.
$$
For this purpose let us show that the kernel of the matrix on the l.h.s. is
empty. As $\bigl((\Tt^E_2)^*\bigr)^{-1}=-\Jj\Tt^E_2\Jj$, we thus have to show
that a vector $\left(\begin{array}{c} v \\ w \end{array}
\right)\in\CC^{2L}$ satisfying
$$
-\,\Jj\;
\left(
\begin{array}{cc}
(T_1T_1^*)^{-1} & {\bf 0} \\
{\bf 0} & {\bf 0}
\end{array}
\right)
\;\Tt^E_2
\;
\left(\begin{array}{c} v \\ w \end{array}
\right)
\;=\;
\Tt^E_2\;\Jj\;
\left(
\begin{array}{cc}
(T_2T_2^*)^{-1} & {\bf 0} \\
{\bf 0} & {\bf 0}
\end{array}
\right)
\;\left(\begin{array}{c} v \\ w \end{array}
\right)
\;,
$$
actually vanishes. Carrying out the matrix multiplications, one readily checks
that this is the case.
\hfill $\Box$

\vspace{.2cm}

Of course, one can regroup the terms in \eqref{eq-help3} into packages of two
successive contributions and each of them is positive by the same argument. If
this bound is uniform for the packages ({\it e.g.} the spectrum of the $T_n$
is uniformly bounded away from $0$), one actually deduces an improved 
lower bound by $C_EN$ for some $C_E>0$.

\subsection{The total rotation number}

In this section, we complete the proof of Theorem~\ref{theo-osci}, in
particular the second part of item (iii). Throughout
$E\in\RR$. The total rotation number is defined by
\begin{equation}
\label{eq-totalphase}
\Theta^E_N
\;=\;
\int^E_{-\infty}de\;
\Im m\;\,\partial_e\;\log(\det(U^e_N)))
\;.
\end{equation}
It will be shown shortly that the integral converges.
Using the notations of Section~\ref{sec-Arnold}, 
$\Theta^E_N=\int^E_\Gamma\omega$ for
$\Gamma=([\Phi^E_N]_\sim)_{E\in\overline{\RR}}$.  
Let $\theta^E_{N,l}$ be the analytic eigenphases of $U^E_N$
as introduced after \eqref{eq-Uasym}.
We deduce after integration of \eqref{eq-totalphase} that 
\begin{equation}
\label{eq-phasesum}
\Theta^E_N
\;=\;
\sum_{l=1}^L\;\bigl(\theta^E_{N,l}-\theta^\Mm_l\bigr)
\;.
\end{equation}
This justifies the term {\sl total rotation number}.  The following result
could also be deduced from the results of the previous section, but its proof
(a homotopy argument) directly 
completes the proof of Theorem~\ref{theo-osci}(iii).

\begin{proposi}
\label{prop-totalphase}
The total rotation number $\Theta^E_N$
is well-defined and satisfies
$$
\lim_{E\to-\,\infty}\Theta^E_N
\;=\;
0
\;,
\qquad
\lim_{E\to\infty}\Theta^E_N
\;=\;
2\pi\,N\,L
\;.
$$
\end{proposi}

\noindent {\bf Proof.} Expanding $\Phi^E_N=\Mm\prod_{n=1}^N\Tt^E_n\,
\Phi_\zeta$ shows
\begin{eqnarray*}
\bigl((\Phi^E_N)^*\Phi^E_N\bigr)^{-1}
& = &
\left(
\Phi_\zeta^*
\,E^{2N}\,
\left(
\begin{array}{cc}
\one & {\bf 0} \\
{\bf 0} & {\bf 0}
\end{array}
\right)^N
\Mm^*\Mm
\left(
\begin{array}{cc}
\one & {\bf 0} \\
{\bf 0} & {\bf 0}
\end{array}
\right)^N
\Phi_\zeta
\;+\;
\Oo(E^{2N-1})
\right)^{-1}
\\
& & 
\\
& = &
E^{-2N}\,
\left(
\begin{array}{c}
\one \\ {\bf 0}
\end{array}
\right)^*\Mm^*\Mm
\left(
\begin{array}{c}
\one \\ {\bf 0}
\end{array}
\right)
\;+\;
\Oo(E^{-2N-1})
\;.
\end{eqnarray*}
Similarly one verifies
$$
(\Phi^E_N)^*\,\Jj\,\partial_E \Phi^E_N
\;=\;
\Oo(E^{2N-2})
\;.
$$
Hence follows
$$
\bigl((\Phi^E_N)^*\Phi^E_N\bigr)^{-1}
\;
(\Phi^E_N)^*\,\Jj\,\partial_E \Phi^E_N
\;=\;
\Oo(E^{-2})
\;,
$$
so that the integral in \eqref{eq-totalphase} exists  
due to \eqref{eq-derivative}, which can alternatively be derived from 
Lemma~\ref{lem-transferwind}.
Furthermore, one deduces from \eqref{eq-Uasym}
that  $\Gamma=([\Phi^E_N]_\sim)_{E\in\overline{\RR}}$ is a closed
path in $\LM_L^\CC$ and its winding number is given by
$$
\langle\, \omega\,|\,\Gamma\,\rangle
\;=\;
\frac{1}{2\pi}\;
\lim_{E\to\infty}\;
\Theta^E_N
\;.
$$
In order to calculate the winding number and prove the last statement of the
proposition, one applies Proposition~\ref{prop-basischange} and
Lemma~\ref{lem-transferwind} iteratively to the path
$\Gamma=([\Mm\prod^N_{n=1}\Tt^E_n\Phi_\zeta]_\sim)_{E\in\overline{\RR}}$.
Alternatively one can use a homotopy $H^N(\lambda)$
from $H^N(1)=H^N_\varphi$ to $H^N(0)$ which is the sum of $L$ 
un-coupled one-dimensional discrete Laplacians (using the
homotopy of the proof of Lemma~\ref{lem-transferwind} on every site $n$). 
For each of the one-dimensional
discrete Laplacians the winding number is again easy to calculate and equal to
$N$.
\hfill $\Box$

\vspace{.2cm}

\noindent {\bf Proof} of  the last statement of Theorem~\ref{theo-osci}(iii).
The homotopy
discussed at the end of the proof of Proposition~\ref{prop-totalphase} is
analytic and hence one deduces $\theta^E_{N,l}-\theta_l^\Mm\to 2\pi N$ for 
$E\to\infty$ for each $l$ as this is the case for the one-dimensional discrete
Laplacian. 
\hfill $\Box$

\vspace{.2cm}

\noindent {\bf Remark} Using Proposition~\ref{prop-totalphase} one can also
give a nice alternative proof of $\partial_E\theta^E_{N,l}\geq 0$.
Indeed, 
according to \eqref{eq-eigencond2}, the unitary $U^E_N$ can be used in order to
calculate the spectrum of $H^N_\varphi$ for every $\varphi\in(0,2\pi)$.
Counting multiplicities, this spectrum consists of $NL$ eigenvalues. 
By Proposition~\ref{prop-totalphase} and
Theorem~\ref{theo-Arnold}, the total number
of passages of eigenvalues $\theta^E_{N,l}$
by $\varphi$ (intersections with the singular cycle $\LM_L^{\varphi}$)
is bounded below by $NL$. As there
cannot be more than $NL$, all these passages have to be in the positive
sense because a passage in the negative sense would lead to at least two more
eigenvalues of $H^N_\varphi$.

\subsection{Telescoping the total rotation number}
 
By the results of the last section and due to the fact that a change of right
boundary condition can change the number of eigenvalues by at most $L$, one
has
\begin{equation}
\label{eq-EVcount}
\left|\;\frac{1}{2\pi}\;\Theta^E_N\,-\,
\#
\left\{
\mbox{ eigenvalues of } H^N_\xi \;\leq\;E\; 
\right\}\;
\right|
\;\leq\;
2\,L
\;.
\end{equation}
Hence $\Theta^E_N$ allows
to count the eigenvalues of $H^N$ up to boundary terms. For this purpose it is
useful to telescope $\Theta^E_N$ into $N$ contributions stemming from the
$L$-dimensional slices:
$$
\Theta^E_N
\;=\;
\sum_{n=1}^{N}\;
\int^E_{-\infty}de\;
\Im m\;\,\partial_e\;
\log
\left(
\frac{\det(U^e_{n})}{
\det(U^e_{n-1})}
\right)
\;.
$$
Here we have used \eqref{eq-Usymdyn} and the fact 
that $U^e_0$ is independent of
$e$. This is indeed a good way to telescope because 
$U^e_{n}=\Cc\,\Mm\,\Tt^e_{n}\,\Mm^{-1}\,\Cc^*\cdot U^e_{n-1}$ so that
Proposition~\ref{prop-basischange} and
Lemma~\ref{lem-transferwind} imply that each summand satisfies
$$
\left|\;
\int^E_{-\infty}de\;
\Im m\;\,\partial_e\;
\log
\left(
\frac{\det(\Cc\,\Mm\,\Tt^e_{n}\,\Mm^{-1}\,\Cc^*\cdot U^e_{n-1})}{
\det(U^e_{n-1})}
\right)
\;\right|
\;\leq\;2\,L
\;.
$$
Moreover, with the notation
\begin{equation}
\label{eq-Centries}
\Cc\,\Mm\,\Tt^E_{n}\,\Mm^{-1}\,\Cc^*
\;=\;
\left(
\begin{array}{cc} A^E_n & B^E_n \\ C^E_n & D^E_n 
\end{array}
\right)
\;,
\end{equation}
the same calculation as in Proposition~\ref{prop-basischange} implies that 
\begin{equation}
\label{eq-telescop}
\Theta^E_N
\;=\;
\sum_{n=1}^{N}\;
\int^E_{-\infty}de\;
\Im m\;\,\partial_e\;
\log
\left(
\,\det\bigl(\,(A^e_{n}\,+\,B^e_{n}\,(U^e_{n-1})^*)
(D^e_{n}\,+\,C^e_{n}\,U^e_{n-1})^{-1}
\;\bigr)\;
\right)
\;.
\end{equation}
Now it is actually possible to apply the fundamental theorem in every summand
by determining the
branch of the logarithm uniquely from the transfer matrix $\Tt^E_n$, and
independent of the prior transfer matrices. 
Indeed, one can factor out $\det(A^e_{n}(D_n^e)^{-1})$ and use the fact that 
$(A^e_{n})^{-1}B^e_{n}$ 
and $(D^e_{n})^{-1}C^e_{n}$ 
have norm less than $1$ by Lemma~\ref{lem-Cayley}, so
that  as in the proof of 
Proposition~\ref{prop-basischange}:
\begin{eqnarray}
\Theta^E_N
& = &
\sum_{n=1}^{N}\;
\Im m\;\Bigl[
\int^E_{-\infty}de\;
\,\partial_e\;
\log
\left(
\,\det(\,A^e_{n}(D_n^e)^{-1}\,)\,
\right)
\nonumber
\\
& &
\label{eq-telescop2}
\\
& &
\;\;\;\;\;+\;
\Tr\left(\log\left(
\one\,+\,(A^E_{n})^{-1}\,B^E_{n}\,(U^E_{n-1})^*
\;\right)\right)
-
\;\Tr\left(\log\left(
\one\,+\,(D^E_{n})^{-1}\,C^E_{n}\,U^E_{n-1}
\;\right)\right)
\Bigr]
\nonumber
\;.
\end{eqnarray}
A refined version of this formula is  
exploited in Section~\ref{sec-covariant}. 

\subsection{The Dean-Martin identity}

In this and the next two 
sections, $\Mm=\one$ and $\xi=\nul$ (Dirichlet boundary
conditions on the right boundary). The Hamiltonian will be denoted by
$H^N$ without a further index. 
According to \eqref{eq-eigencond},  $E$ is
an eigenvalue of $H^N$  if and only if 
$$
\det
\left(
\bigl( \nul\;\,\one\bigr)\;\Jj\;\Phi^E_N
\,\right)
\;=\;
\det
\left(
\bigl( \one\;\nul\bigr)\;
\prod_{n=1}^N\Tt_n^E\;
\Phi_\zeta\;
\right)
\;=\;
0\;.
$$
Due to \eqref{eq-eigencond},
the multiplicity of the zero is the multiplicity of the
eigenvalue. Therefore the l.h.s. of this equation is a polynomial of degree
$NL$ in $E$ with zeros exactly at the $NL$ eigenvalues of
$H^N$. 
Comparing the leading order coefficient, one deduces a formula for the
characteristic polynomial:
\begin{equation}
\label{eq-charid1}
\det(E\,\one-H^N)
\;=
\;
\det
\left(
\bigl( \one\;\nul\bigr)\;
\Phi^E_N\,
\right)
\;\;\prod_{n=1}^N\det(T_n^{-1})
\;.
\end{equation}
In order to find a recurrence relation for the characteristic polynomials, let
us suppose that $\Im m(E)>0$ and note that $Z_N^E=
\bigl(\one\;\nul\bigr)\,\Phi_N^E
\left(\bigl(\one\;\nul\bigr)\,\Phi_{N-1}^E \right)^{-1}T_N$.
Taking the determinant of this formula, the identity
\eqref{eq-charid1} applied twice gives 
\begin{equation}
\label{eq-DM}
\det(Z^E_N)
\;=\;
\frac{\det\bigl(E\,\one-H^N\bigr)}{
\det\bigl(E\,\one- H^{N-1}\bigr)}
\;.
\end{equation}
Let us call this the Dean-Martin identity, due to the contribution
\cite{DM}. 
%
%
%
%
These authors then used the identity \eqref{eq-DM} at real energies 
in order to calculate the spectrum of
$H^{N-1}$ by counting the singularities of $\det(Z^E_N)$. This can be made
more explicit by adding a small imaginary part $\delta>0$ to the energy. Then
consider the path  $E\in\RR\mapsto\det(Z^{E+\imath\delta}_N)\in\CC$. 
Even though $Z^{E+\imath\delta}_N$ is in the upper half-plane, its
determinant may well have a negative imaginary part. 
However, it never takes the values
$\pm\,\imath$. Now each passage of the path (near) by an eigenvalue of 
$H^{N-1}$ leads to an arc in either the upper or lower half-plane 
with passage by either $\imath /\delta$
or $-\imath/\delta$,
pending on the sign of the numerator in \eqref{eq-DM}. 
Both arcs turn out to be
in the positive orientation. A multiple eigenvalue
leads to a multiple arc. The topologically interesting quantity is the winding
numbers of the path around $\imath$ and $-\imath$. Calculating the sum of the
corresponding phase integrals 
gives a total rotation number which in the limit $\delta\to 0$
coincides with $\Theta^E_N$. This allows to give a nice alternative, but
considerably more complicated proof of 
Proposition~\ref{prop-totalphase}. In the case $L=1$, all the arcs are in the
upper half-plane and the argument just sketched is particularly simple.

\subsection{Green's function and continued fraction expansion}
\label{sec-Green}

The aim of this short section is to illustrate the use of the dynamics in the 
upper half-plane. 
For $\Im m (E)>0$, the $L\times L$ 
Green's matrix $G^{E,N}_{n,m}$ for
$1\leq n,m\leq N$ is defined by
$$
G^{E,N}_{n,m}(k,l)
\;=\;
\langle n,k|\,(H^N-E\,\one)^{-1}\,|m,l\rangle
\;,
\qquad
k,l=1,\ldots,L\;.
$$
It follows from
the Schur complement formula that
$$
G_{N,N}^{E,N}
\;=\;
\left(
V_N-E\,\one-T_N G^{E,N-1}_{N-1,N-1}T_N
\right)^{-1}
\;.
$$
Iteration of this formula 
gives a matricial continued fraction expansion:
$$
G_{N,N}^{E,N}
\;=\;
\left(
V_N-E\,\one-T_N
\left(
\cdots
\left(
V_2-E\,\one
-
T_2(V_1- E\,\one+\zeta)^{-1}T_2
\right)^{-1}
\cdots
\right)^{-1}T_N
\right)^{-1}
\;.
$$
As $Z_1^E=E\,\one-V_1-\zeta$, one sees that this is just the iteration of
the Ricatti equation. Hence one deduces
$$
G_{N,N}^{E,N}
\;=\;
-\,
(Z^E_N)^{-1}
\;=\;
-\,\pi(\Phi^E_{N})^{-1}
\;,
$$
where the inverse $(Z^E_N)^{-1}$ exists because it
is given by the M\"obius transformation
of $Z^E_N\in\UM_L^\CC$ 
with $\imath\,\Jj\,\in\,$SP$(2L,\CC)$. Hence $G_{N,N}^{E,N}\in
\UM_L^\CC$. Furthermore the geometric resolvent identity shows
$$
G_{1,N}^{E,N}
\;=\; 
-\,
G_{1,N-1}^{E,N-1}
\,T_N\,
G_{N,N}^{E,N}
\;.
$$
%
%
These identities allow us to note a few
useful identities linking the entries of the transfer matrix
$$ 
\Tt^E_N\cdot\ldots\cdot \Tt_1^E
\;=\;
\left(
\begin{array}{cc} A & B \\ C & D 
\end{array}
\right)
\;,
$$
to the Green's function:
$$
A^{-1}\;=\;-\,G^{E,N}_{1,N}\;,
\qquad
C^{-1}\;=\;-\,G^{E,N-1}_{1,N-1}\,T_N\;,
\qquad
CA^{-1}\;=\;-\,G^{E,N}_{N,N}\;,
\qquad
A^{-1}B\;=\;G^{E,N}_{1,1}\;.
$$
%
%
It is also possible to express $B^{-1}$, $D^{-1}$, $DB^{-1}$ and
$C^{-1}D$ in terms of Green's functions.

\subsection{Eigenvalue interlacing}

In this section, the above information on the spectrum of Jacobi matrices with
matrix entries is complemented by a
simple consequence of the min-max principle. 
In the case $L=1$ of a Jacobi
matrix, this is the
theorem on alternation of zeros of the associated 
orthogonal polynomials. It also implies
that the bottom (resp. top) of the spectrum of $H^N$ is less (resp. larger) 
than or equal the bottom (resp. top) of the spectrum of $H^{N-1}$.

\begin{proposi}
\label{prop-minmax}
Let $H^N$ and $H^{N-1}$ be defined with Dirichlet boundary conditions on the
right boundary.
Then the eigenvalues of
$H^N$ and $H^{N-1}$ satisfy the following interlacing property:
$$
E^N_j
\;\leq\;
E^{N-1}_j
\;\leq\;
E^N_{j+L}
\;,
\qquad
j=1,\ldots,(N-1)L\;.
$$
\end{proposi}

\noindent {\bf Proof.}
Let $\Hh_N=\ell^2(1,\ldots,N)\otimes \CC^L$ and $\Pi_N$ the projection in
$\Hh_N$ on the states on the right boundary, namely, in Dirac notation, on the
span of $(|N,l\rangle)_{l=1,\ldots,L}$. Hence $\Hh_{N-1}\cong (\one -\Pi_N)
\Hh_N$ and $\Hh_{N-1}\subset\Hh_N$ with the natural embedding. Also
$H^{N-1}|\psi\rangle = H^N|\psi\rangle$ for $\psi \in\Hh_{N-1}$, {\it i.e.}
$\Pi_N\psi=0$ (the natural embedding is suppressed in this notation). 
The min-max principle states:
$$
E^N_j
\;=\;
\sup_{U\subset \Hh_N\,,\;\dim(U)\leq j}
\;\;
\inf_{\psi\in U^\perp\,,\;\|\psi\|=1} \;\;
\langle \psi\,|\,H^N\,|\,\psi\rangle
\;,
$$
where the supremum is over subspaces $U$ of $\Hh_N$, and the infimum over
unit vectors in their orthogonal complement. For $H^{N-1}$, the above facts 
imply
$$
E^{N-1}_j
\;=\;
\sup_{U\subset \Hh_N\,,\;\dim(U)\leq j\,,\;\Pi_NU=0}
\;\;\;\;
\inf_{\psi\in U^\perp\,,\;\|\psi\|=1\,,\;\Pi_N\psi=0} \;\;
\langle \psi\,|\,H^N\,|\,\psi\rangle
\;,
$$
where the orthogonal complement is calculated in $\Hh_N$.
Hence the inequality $E^{N-1}_j\geq E^{N}_j$ follows because the condition 
$\Pi_NU=0$ is redundant and then one obtains a lower bound 
by dropping the constraint $\Pi_N\psi=0$. Next, for a subspace
$U\subset \Hh_N$ with $\Pi_NU=0$,
let $\tilde{U}=U\oplus \Pi_N \Hh_N$. Then
$\dim(\tilde{U})=\dim(U)+L$. Furthermore, the conditions 
$ \psi\in U^\perp$ and $\Pi_N\psi=0$
are equivalent to $ \psi\in \tilde{U}^\perp$. 
Therefore, upon relaxing the constraints on the supremum:
$$
E^{N-1}_j
\;\geq\;
\sup_{\tilde{U}\subset \Hh_N\,,\;\dim(\tilde{U})\leq j+L}
\;\;\;\;
\inf_{\psi\in \tilde{U}^\perp\,,\;\|\psi\|=1} \;\;
\langle \psi\,|\,H^N\,|\,\psi\rangle
\;=\;
E^N_{j+L}\;,
$$
which is the second inequality.
\hfill $\Box$

\section{Jacobi matrices with random matrix entries}
\label{sec-covariant}

In this section we consider Jacobi matrices $H^N(\omega)$ with matrix entries
$\omega=(V_n,T_n)_{n\geq 1}$ which are independent and identically distributed
random variables drawn from a bounded ensemble
$(V_\sigma,T_\sigma)_{\sigma\in\Sigma}$ of symmetric and positive real 
matrices. Expectation w.r.t. to their distribution will be denoted
by  $\EE_\sigma$ or simply by $\EE$. All formulas in 
Section~\ref{sec-DOSformula}
also hold for
more general covariant operator families and systems without time-reversal
symmetry. 
Associated to each $\omega$ 
are transfer matrices $\Tt^E_n(\omega)$,
Lagrangian planes $\Phi^E_n(\omega)$, their parame\-trizations 
$Z^E_n(\omega)$ and
$U^E_n(\omega)$, matrix entries $A^E_n(\omega)$ and $B^E_n(\omega)$ as in
\eqref{eq-Centries}, total rotations, etc.
In order not to overload notation, the index $\omega$ is suppressed
throughout. The basis change $\Mm$ will be taken independent of $\omega$
though.

\subsection{Integrated density of states and sum of Lyapunov exponents}
\label{sec-DOSformula}

The integrated density of states of a random family of 
Jacobi matrices with matrix entries is defined
by \cite{CS,BL,CL}
$$
\Nn(E)
\;=\;
\lim_{N\to\infty}
\;\frac{1}{N\,L}\;\EE\;
\#
\left\{
\mbox{ eigenvalues of } H^N \;\leq\;E\; 
\right\}
\;.
$$
According to \eqref{eq-EVcount}, \eqref{eq-telescop} and the fact that
$C_e=\overline{B_e}$ and $D_e=\overline{A_e}$, one has
\begin{equation}
\label{eq-IDS}
\Nn(E)
\;=\;
\lim_{N\to\infty}
\;\frac{1}{N\,L}
\;\sum_{n=1}^N\;
\int^E_{-\infty}\frac{de}{\pi}\;
\Im m\;\,\partial_e\;
\EE\;\log
\left(
\,\det\bigl(\,A^e_{n}\,+\,B^e_n\, (U^e_{n-1})^*\,\bigr)\,
\right)
\;.
\end{equation}
For a fixed energy, this quantity can also be understood as a rotation
number in the sense of Ruelle \cite{Rue}.
The second ergodic quantity considered here is the averaged sum of the positive
Lyapunov exponents, denoted shortly by $\gamma(E)$ here. For any complex
energy $E\in\CC$, it can be defined by \cite{CS,BL,KS,CL,SB}
\begin{equation}
\label{eq-Lyapdef}
\gamma(E)
\;=\;
\lim_{N\to\infty}
\;\frac{1}{N\,L}\;\EE\;
\log
\left(
\;\left\|
\Lambda^L\left(\prod^N_{n=1}\Tt^E_n  \right)
\right\|\;
\right)
\;,
\end{equation}
where $\Lambda^L\Tt$ is the $L$-fold exterior product (second quantization as
for evolution group) of the
symplectic matrix $\Tt$, and the norm denotes the operator norm on
the fermionic Fock space $\Lambda^L\CC^{2L}$. 
It is well known that $\gamma(E)$ is subharmonic in $E$ 
\cite{CS,CL}. Furthermore,
the Thouless formula linking $\Nn$ and $\gamma$ holds \cite{CS,KS}. 
Actually this is the integrated version of the Kramers-Kr\"onig relation
stating that
$\gamma(E)+\imath\,\pi\,\Nn(E)$ for real $E$ is the 
boundary value of a Herglotz
function (which is given by the expectation value of the trace of the logarithm
of the Weyl-Titchmarch matrix \cite{KS}, for which one has a
matrix-valued Herglotz representation \cite{GT}). 
This is reflected by the following proposition showing that $\gamma$
and $\pi\,\Nn$ can be calculated as real and imaginary part of the Birkhoff sum
associated to a single complex valued additive cocycle.

\begin{proposi}
\label{prop-lyap}
For $E$ with $\Im m(E)\geq 0$,
$$
\gamma(E)
\;=\;
\lim_{N\to\infty}
\;\frac{1}{N\,L}
\;\sum_{n=1}^N\;
\Re e\;
\EE\;\log
\left(
\,\det\bigl(\,A^E_{n}\,+\,B^E_n\, (U^E_{n-1})^*\,\bigr)\,
\right)
\;.
$$
\end{proposi}

\noindent {\bf Proof.}
Clearly one may replace $\Tt^E_n$ in  \eqref{eq-Lyapdef} by
$\Mm\Tt^E_n\Mm^{-1}$ because the boundary contributions drop out in the limit.
Furthermore, instead of calculating the
operator norm in \eqref{eq-Lyapdef}, one may insert a real Lagrangian plane 
$\Phi_0=(\phi_1,\ldots,\phi_L)$ as initial conditions
$$
\gamma(E)
\;=\;
\lim_{N\to\infty}
\;\frac{1}{N\,L}\;\EE\;
\log
\left(
\;\left\|
\Lambda^L\left(\prod^N_{n=1}\Mm\,\Tt^E_n\,\Mm^{-1}  \right)\;
\phi_1\wedge\ldots\wedge\phi_L
\right\|\;
\right)
\;,
$$
where now the norm is that of a vector in $\Lambda^L\CC^{2L}$
\cite[A.III.3.4]{BL} (for covariant, but not necessarily 
random Jacobi matrices
with matrix entries, this holds as long as $\EE$ contains an average 
over $\Phi_0$ w.r.t. some
continuous measure \cite{JSS,SB}).  Recalling that the
norm in $\Lambda^L\CC^{2L}$ is calculated with the determinant, it follows
that
$$
\gamma(E)
\;=\;
\lim_{N\to\infty}
\;\frac{1}{2\,N\,L}\;\EE\;
\log
\left(\;\det\left(
\;\Phi_0^* \,\Bigl(\prod^N_{n=1}\Mm\,\Tt^E_n\,\Mm^{-1} \Bigr)^*
\;
\Bigl(\prod^N_{n=1}\Mm\,\Tt^E_n\,\Mm^{-1}  \Bigr)\,\Phi_0
\,\right)\;\right)
\;.
$$
Now $\Bigl(\prod^N_{n=1}\Mm\,\Tt^E_n\,\Mm^{-1}  \Bigr)\,\Phi_0=\Phi^E_N$ and
one may telescope (boundary terms vanish in the limit) 
and insert the Cayley transformation: 
$$
\gamma(E)
\;=\;
\lim_{N\to\infty}
\;\frac{1}{2\,N\,L}\;\EE\;\sum_{n=1}^N\;
\log
\left(\frac{\det\bigl((\Cc\Phi^E_n)^*\,(\Cc\Phi^E_n)\bigr)}{
\det\bigl((\Cc\Phi^E_{n-1})^*\,(\Cc\Phi^E_{n-1})\bigr)}
\right)
\;.
$$
In each term, one can now apply Lemma~\ref{lem-RNcocycle} 
for $\Tt=\Cc\Mm\Tt^E_n\Mm^{-1}\Cc^*$ and $\Phi=\Cc\Phi^E_{n-1}$.
The hypothesis of the lemma 
are indeed satisfied for any $E$ with $\Im m(E)\geq 0$ 
because of the arguments
in Section~\ref{sec-Moebius}. According to the
definition of $U^E_n$, it therefore follows
$$
\gamma(E)
\;=\;
\lim_{N\to\infty}
\;\frac{1}{2\,N\,L}\;\EE\;\sum_{n=1}^N\;
\log
\left(
\;
\frac{\det\bigl(\,(U^E_n)^*\,U^E_n+\one\,\bigr)}{
\det\bigl(\,(U^E_{n-1})^*\,U^E_{n-1}+\one\,\bigr)}
\;\,
\left|\;
\det\bigl(\,A^E_{n}\,+\,B^E_n\, (U^E_{n-1})^*\,\bigr)
\,\right|^2
\;\right)
\;.
$$
The first contribution telescopes back again and the boundary term at $N$ is
bounded because $1\leq\det(U^*U+\one)\leq 2^L$ for every
$U\in\DM_L^\CC\cup\partial_L\DM_L^\CC$. 
Hence the first contribution vanishes in the
limit. 
The second contribution is precisely the term appearing in the proposition.
\hfill $\Box$


\subsection{Random perturbations}
\label{sec-perturb}

This section gives the precise hypothesis of Theorem~\ref{theo-perturb} and
then provides the proof. Hence $V_n$
and $T_n$ are random and depend on a coupling parameter $\lambda\geq 0$ 
as described in the introduction and they give rise to transfer
matrices $\Tt^E_n(\lambda)$ which depend analytically on $\lambda$ (lower
regularity is actually sufficient). Throughout this section $E\in\RR$. 
The first step of the analysis consists in the
symplectic diagonalization of $\Tt^E=\Tt^E_n(0)$ 
by an adequate symplectic basis change $\Mm$:
\begin{equation}
\label{eq-diagonalized}
\Mm\,\Tt^E_n(\lambda)\,\Mm^{-1}
\;=\;
\Rr\;\exp\bigl(\,\lambda\,\Pp_n\,+\,\Oo(\lambda^2)\,\bigr)
\;.
\end{equation}
Here (and in matrix equations below) the expression 
$\Oo(\lambda^2)$ means that we have an
operator norm estimate on the remainder. Furthermore $\Pp_n$ is a (random)
element of the Lie algebra sp$(2L,\RR)$ calculated from $(v_n,t_n)$  and
$\Rr$ is of the  symplectic normal form of the free transfer matrix $\Tt^E$
chosen as follows. The eigenvalues of $\Tt^E$ form quadruples
$(\lambda,1/\lambda, \overline{\lambda},1/\overline{\lambda})$ which collapse
to pairs $(\lambda,1/\lambda)$ if $\lambda\in S^1$ and $\lambda\in\RR$. If
$\lambda\in S^1$, one speaks of an elliptic channel. Let there be $L_e$ of
them, denote their eigenvalues by
$e^{\imath \eta_1},\ldots,e^{\imath\eta_{L_e}}$ and set
$\eta=\mbox{diag}(\eta_1,\ldots,\eta_{L_e})$. As $\Tt^E$ is supposed to be
diagonalizable, the remaining $L_h=L-L_e$ channels are hyperbolic. The moduli
of their eigenvalues are $e^{\kappa_l},e^{-\kappa_l}$, with $\kappa_l>0$ 
and for
$l=1,\ldots,L_h$. Set $\kappa=\mbox{diag}(\kappa_1,\ldots,\kappa_{L_h})$.
If a hyperbolic channel stems from a quadruple, it moreover
contains a rotation by the phase of its eigenvalue $\lambda$. This will be
described by $S\in\,$O$(L_h)$ which is a tridiaganol orthogonal matrix
containing only either $1$ or $2\times 2$ rotation matrices
on the diagonal and which satisfies $[S,e^\kappa]=0$. The symplectic basis
change $\Mm$ is then chosen such that
$$
\Rr
\;=\;
\left(
\begin{array}{cccc}
S\,e^\kappa & \nul          &  \nul           &  \nul       \\
\nul        & \cos(\eta)    &  \nul           &  \sin(\eta) \\
\nul        & \nul          &  S\,e^{-\kappa} &  \nul       \\
\nul        & -\,\sin(\eta) &  \nul           &  \cos(\eta)
\end{array}
\right)\;,
$$
Furthermore, let $P_h$ and $P_e$ denote
the projections ($L\times L$ matrices) onto the hyperbolic and elliptic
channels. In particular, $P_h+P_e=\one$ and 
$\mbox{diag}(P_h,P_h)$ as well as $\mbox{diag}(P_e,P_e)$
commute with $\Rr$. The reader may consult \cite{SB} where the basis change
$\Mm$ is constructed explicitly 
for the example of the Anderson model on a strip. 
Next let us state the precise hypothesis
of Theorem~\ref{theo-perturb}.

\vspace{.2cm}

\noindent {\bf Hypothesis:} {\sl The expansion factors $\kappa_l$ and rotation
phases $\eta_l$ satisfy}
$$
g_h\;=\;\min_{1\leq l\leq L_h}\;(1-e^{-\kappa_l})\;>\;0
\;,
\qquad
g_e\;=\;\min_{1\leq l,k\leq L_e}\;
\bigl|1-e^{\imath(\eta_l+\eta_k)}\bigr|\;>\;0
\;.
$$

\vspace{.2cm}

In order to develop the perturbation theory, some further notations are 
needed: 
$$
\Cc\,\Mm\,\Tt^E_{n}(\lambda)\,\Mm^{-1}\,\Cc^*
\;=\;
\left(
\begin{array}{cc} A^E_n(\lambda) & B^E_n(\lambda) \\ 
\overline{B^E_n(\lambda)} & \overline{A^E_n(\lambda)} 
\end{array}
\right)
\;=\;
\left(
\begin{array}{cc} A+\lambda a_n & B+\lambda b_n \\ 
\overline{B+\lambda b_n} & \overline{A+\lambda a_n} 
\end{array}
\right)
\;+\;\Oo(\lambda^2)
\;.
$$
Comparing with \eqref{eq-diagonalized}, one checks that
$$
A
\;=\;
\left(
\begin{array}{cc} S\,\cosh(\kappa) & \nul \\ 
\nul & e^{\imath\eta} 
\end{array}
\right)
\;,
\qquad
B
\;=\;
\left(
\begin{array}{cc} S\,\sinh(\kappa) & \nul \\ 
\nul & \nul
\end{array}
\right)
\;,\qquad
A\,\pm\,B
\;=\;
\left(
\begin{array}{cc} e^{\pm \kappa} & \nul \\ 
\nul & e^{\imath\eta} 
\end{array}
\right)
\;,
$$
and
$$
a_n
\;=\;
\left(
\begin{array}{c} A \\ B
\end{array}
\right)^t
\Cc\,\Pp_n\,\Cc^*
\left(
\begin{array}{c} \one \\ \nul
\end{array}
\right)
\;,
\qquad
b_n
\;=\;
\left(
\begin{array}{c} A \\ B
\end{array}
\right)^t
\Cc\,\Pp_n\,\Cc^*
\left(
\begin{array}{c} \nul \\ \one
\end{array}
\right)
\;.
$$
Note that both $A$ and $B$ commute with both $P_h$ and $P_e$.

\vspace{.2cm}

According to Section~\ref{sec-DOSformula}, the averaged 
Lyapunov exponent and IDS at a real
energy $E$ for the $\lambda$-dependent random operators $H^N(\lambda)$ are 
given by
$$
\gamma_\lambda(E) 
\;+\;\imath\,\pi\,\Nn_\lambda(E)
\;=\;
\lim_{N\to\infty}
\;\frac{1}{N\,L}
\;\sum_{n=1}^N\;
\int^E_{-\infty}{de}\;
\partial_e\;
\EE\;\log
\left(
\,\det\bigl(\,A^e_{n}(\lambda)\,+\,B^e_n(\lambda)\, 
(U^e_{n-1}(\lambda))^*\,\bigr)\,
\right)
\;.
$$
Let us expand the integrand 
\begin{equation}
\label{eq-exhelp0}
\log
\left(
\,\det\bigl(\,A^E_{n}(\lambda)\,+\,B^E_n(\lambda)\, 
(U^E_{n-1}(\lambda))^*\,\bigr)\,
\right)
\;=\;
\log
\left(
\,\det\bigl(\,A^E_{n}(\lambda)\,+\,B^E_n(\lambda)\,P_h 
\,\bigr)\,
\right)
\;+\;
J_n(\lambda)
\;,
\end{equation}
where
\begin{equation}
\label{eq-exhelp}
J_n(\lambda)
\;=\;
\Tr\left(
\log
\bigl(\,\one\;+\,
(A^E_{n}(\lambda)\,+\,B^E_n(\lambda)\,P_h )^{-1}
B^E_{n}(\lambda)\,
(U^E_{n-1}(\lambda)-P_h)^*
\,\bigr)\,
\right)
\;,
\end{equation}
which is possible because
$A^E_{n}(\lambda)\,+\,B^E_n(\lambda)\,P_h=A+B+\Oo(\lambda)$ is invertible for
$\lambda$ sufficiently small. Note that only $J_n(\lambda)$ depends on
$U^E_{n-1}$, while the first contribution gives an contribution to IDS and
Lyapunov exponent which can be readily calculated, similar as in
\eqref{eq-telescop2}.  Hence we need to focus on
the control of Birkhoff averages of $J_n(\lambda)$.  
For this purpose, one would first like to expand
the logarithm in \eqref{eq-exhelp}. As $B^E_n(\lambda)=BP_h+\Oo(\lambda)$, 
one therefore
has to show that $P_hU^E_n(\lambda)-P_h$ is small in norm. This means that the
hyperbolic part of the dynamics (at $\lambda=0$) 
alines $P_hU^E_n(\lambda)$ deterministically with $P_h$ 
up to small corrections due to the random perturbation. The following lemma is
a strengthening of prior results \cite{SB}
on this dynamical separation of hyperbolic
and elliptic channels.

\begin{lemma}
\label{lem-hyp}
Let $U^E_0=\one$ {\rm (}choice of initial condition{\rm )}. Then there exist
positive constants $c_1,c_2$ such that for $\lambda<\frac{g_h^2}{4c_1c_2}$ and
all $n\geq 1$, one has
$$
\|\,
P_hU^E_n(\lambda)-P_h
 \,\|
\;\leq\;
\frac{2\,c_1\,\lambda}{g_h}\;.
$$
\end{lemma}

\noindent {\bf Proof.} 
For sake of notational simplicity, let $P$ denote $P_h$ within this proof.
Let $U$ be a symmetric unitary and set 
$U'=\Cc\Mm\Tt^E_n(\lambda)\Mm^{-1}\Cc^*\cdot U$. 
We will show that uniformly in $n$
holds 
\begin{equation}
\label{eq-exhelp2}
\|\,
PU'-P
 \,\|
\;\leq\;
(1-g_h)\;
\|\,
PU-P
 \,\|
\;+\;c_1\,\lambda
\;+\;c_2\,
\|\,
PU-P
 \,\|^2
\;.
\end{equation}
An elementary dynamical argument then allows to conclude the proof (the
constants $c_1$ and $c_2$ are then the same as in the statement of the lemma). 
In order to prove 
\eqref{eq-exhelp2}, let us first note that
$\lambda\mapsto\Cc\Mm\Tt^E_n(\lambda)\Mm^{-1}\Cc^*\cdot U$ is an analytic 
path of unitaries. Hence the eigenvalues vary analytically in $\lambda$
\cite{Kat}. Therefore $U'=\Cc\Rr\Cc^*\cdot U+R_1$ where $R_1$ depends on $U$
and $\Tt^E_n(\lambda)$, but one has the norm bound $\|R_1\|\leq c_1\lambda$
uniformly in $n$ and $U$ (because of the uniform bounds on the norms of
$\Pp_n$ and the error terms in \eqref{eq-diagonalized}). Therefore it is
sufficient to show \eqref{eq-exhelp2} for $\lambda=0$. 

\vspace{.1cm}

By definition of the M\"obius action, $U'=(AU+B)(BU+\overline{A})^{-1}$ so
that 
$$
PU'
\;=\;
P\bigl(
\,A(PU-P)\,+\,(A+B)\,\bigr)
\,(\,\overline{A}+B)^{-1}\,
\bigl(\one\,+
\,B\,(PU-P)\,(\,\overline{A}+B)^{-1}\,\bigr)^{-1}
\;.
$$
Now one can expand the last inverse in $(PU-P)$ to first order, 
with an error term bounded by 
$\|PU-P\|^2$. Then multiplying out all the remaining factors shows that
\begin{equation}
\label{eq-exhelp3}
PU'-P
\;=\;
P\,(A-B)\,(PU-P)\,(\overline{A}+B)^{-1}
\;+\;R_2
\;,
\end{equation}
with an error term that satisfies $\|R_2\|\leq c_2 \,\|PU-P\|^2$. 
Now $\| (\overline{A}+B)^{-1} \|=1$ 
and $\|P(A-B)\|=\max_{1\leq l\leq L_h}\;e^{-\kappa_l}=1-g_h$ which
implies \eqref{eq-exhelp2} for $\lambda=0$.
\hfill $\Box$

\vspace{.2cm}

Now it is possible to expand the logarithm in \eqref{eq-exhelp} because
$B^E_{n}(\lambda)P_e=\Oo(\lambda)$ so that 
Lemma~\ref{lem-hyp} implies
$B^E_{n}(\lambda)(U^E_{n-1}(\lambda)-P_h)^*=\Oo(\lambda)$ (here still all
error terms are norm bounded). Hence
$$
J_n(\lambda)
\;=\;
\Tr\left(
\,
(A^E_{n}(\lambda)\,+\,B^E_n(\lambda)\,P_h )^{-1}
B^E_{n}(\lambda)\,
(U^E_{n-1}(\lambda)-P_h)^*
\,
\right)
\;+\;\Oo\Bigl(\frac{L\lambda^2}{g_h^2}\Bigr)
\;,
$$
where the $L$ comes from carrying out the trace after having applied the norm
bound. Expanding $A^E_{n}(\lambda)$ and $B^E_{n}(\lambda)$, using the
commutativity of $A$ and $B$ with $P_h$ and $P_e$ and invoking
Lemma~\ref{lem-hyp} in order 
to show  $P_hU^E_{n}(\lambda)P_e=\Oo(\lambda/g_h)$ 
now implies
$$
J_n(\lambda)
\;=\;
\Tr\Bigl(
\,
(A+B )^{-1}
\bigl(
(P_hU^E_{n-1}(\lambda)^*P_h-P_h)
\;+\;
\lambda\;
b_{n}\,
P_eU^E_{n-1}(\lambda)^*P_e 
\bigr)
\,
\Bigr)
\;+
\;\Oo\Bigl(\frac{L\lambda^2}{g_h^2}\Bigr)
\;.
$$
Setting
$$
I_h(N)
\;=\;
\EE\;
\frac{1}{N\,L}
\;
\sum_{n=0}^{N-1}\;
(P_hU^E_n(\lambda)P_h\,-\,P_h)
\;,
\qquad
I_e(N)
\;=\;
\EE\;
\frac{1}{N\,L}
\;
\sum_{n=0}^{N-1}\;
P_e\,U^E_n(\lambda)P_e
\;,
$$
one has
$$
\EE\;
\frac{1}{N\,L}
\;
\sum_{n=1}^N\;J_n(\lambda)
\;=\;
\Tr\left(
\,
(A+B )^{-1}
I_h(N)^*\,\right)
\;+\;\lambda\,
\Tr\left(
\,
(A+B )^{-1}\EE_\sigma(b_\sigma)\,
I_e(N)^*\,\right)
\;+\;
\Oo\Bigl(\frac{\lambda^2}{g_h^2}\Bigr)
\;.
$$

In order to calculate and bound the two traces, we
will use the Hilbert-Schmidt spaces $\Hh_h$ and $\Hh_e$ of complex matrices
respectively of size 
$L_h\times L_h$ and $L_e\times L_e$, furnished with the scalar
product $\langle C|D\rangle_2=\Tr(C^*D)$. The corresponding norms are 
$\|C\|_2=\Tr(C^*C)^{\frac{1}{2}}$. They satisfy 
norm the inequality $\|C\|_2\leq \sqrt{L}\; \|C\|$ w.r.t. to the operator
norm (where one may, of course,
also use respectively $L_h$ and $L_e$ instead of $L$). For a $L\times L$
matrix $C$, we will identify $P_hCP_h$ and $P_eCP_e$ with vectors in
respectively $\Hh_h$ and $\Hh_e$. Let us first focus on the second trace in
the last expression.
The Cauchy-Schwarz inequality implies
$$
\Tr\left(
\,
(A+B )^{-1}\,\EE_\sigma(b_\sigma)\,
I_e(N)^*\,\right)
\;\leq\;
\sqrt{L}\;\|P_e(A+B )^{-1}\,\EE_\sigma(b_\sigma) P_e\|\;\|I_e(N)\|_2
\;.\;
$$
As the operator norm appearing on the r.h.s. is bounded,
the following lemma shows that this trace is of order $\Oo(\lambda/\sqrt{L})$ 
and hence
does not contribute to leading order.

\begin{lemma}
\label{lem-oscisum}
There exist positive constants $c_1$ and $c_2$ such that
$$
\|I_e(N)\|_2
\;\leq\;
\frac{1}{g_e\,\sqrt{L}}\;
\left(c_1\,\lambda\;+\;c_2\,\frac{1}{N}\right)
\;.
$$
\end{lemma}

\noindent {\bf Proof.}
This is a matrix version of the oscillatory sum argument in \cite{PF,SB}.
First note that for each summand in $I_e(N)$, one has   
$P_e\,U^E_n(\lambda)P_e=e^{\imath\eta}P_e\,U^E_{n-1}(\lambda)P_e
e^{\imath\eta}+\Oo(\lambda)$. Thus
$$
I_e(N)
\;=\;
e^{\imath\eta}\,I_e(N)\,e^{\imath\eta}
\;+\;R_1
\;+\;R_2
\;,
$$
with an average error term $R_1$ satisfying $\|R_1\|\leq c_1\,\lambda/L$ and
boundary terms $R_2$ satisfying $\|R_2\|\leq c_2/(NL)$. Now let us define the
super-operator $D_\eta:\Hh_e\to\Hh_e$ by $D_\eta(C)=
e^{\imath\eta}Ce^{\imath\eta}$. This operator is diagonal and 
the hypothesis $g_e>0$ implies that
$(\one-D_\eta)^{-1}$ exists and its norm is bounded by $1/g_e$. 
As $I_e(N)=(\one-D_\eta)^{-1}(R_1+R_2)$, it follows that
$\|I_e(N)\|_2\leq (\|R_1\|_2+\|R_2\|_2)/g_e$ which leads to the desired bound.
\hfill $\Box$

\vspace{.2cm}

A similar argument allows to calculate the remaining trace.

\begin{lemma}
\label{lem-hypsum}
One has
$$
\Tr\left(
\,
(A+B )^{-1}
I_h(N)^*\,\right)
\;=\;
2\,\lambda\;\Re e\;
\Tr\left(
\,
(e^{2\kappa}-\one )^{-1}P_h\;\EE_\sigma(a_\sigma+b_\sigma)\;P_h
\,\right)
\;+\;
\Oo\left(
\frac{\lambda^2}{g_h^3}\,,\,
\frac{1}{g_h\,N}\right)
\;.
$$
\end{lemma}

\noindent {\bf Proof.} One first has to refine \eqref{eq-exhelp3} and include
the $\Oo(\lambda)$ contribution. 
Invoking Lemma~\ref{lem-hyp} at several reprises, some lengthy
but straightforward algebra shows
$$
P_hU^E_n(\lambda)P_h-P_h
\;=\;
S\,e^{-\kappa}\;(P_hU^E_{n-1}(\lambda)P_h-P_h)\,e^{-\kappa}\,S^t
\;+\;
2\,\lambda\;\Re e\;
P_h\;\EE_\sigma(a_\sigma+b_\sigma)\;P_h\,e^{-\kappa}\,S^t
\;+\;
L\,R_1
\;,
$$
with $\|R_1\|\leq c_1\,\lambda^2/(Lg_h^2)$ and the formula is understood as
identity for operators on $\Hh_h$.
Now define the super-operator $D_\kappa:\Hh_h\to\Hh_h$ by $D_\kappa(C)=
Se^{-\kappa}Ce^{-\kappa}S^t$. One directly checks 
$\|(\one-D_\kappa)^{-1}\|\leq 1/g_h$. As
above, 
$$
I_h(N)
\;=\;
(\one-D_\kappa)^{-1}
\Bigl(\frac{2\,\lambda}{L}\;\Re e\;
P_h\;\EE_\sigma(a_\sigma+b_\sigma)\;P_h\,e^{-\kappa}\,S^t
\;+\;
R_1
\;+\;
R_2\Bigr)
\;,
$$
where $\|R_2\|\leq c_2/(NL)$.
Replacing this into the trace and bounding the error terms by the
Cauchy-Schwarz inequality just as before Lemma~\ref{lem-oscisum} allows to
bound the error terms. The leading order contribution can be calculated using
the identity
$$
S^t\,e^{-\kappa}\;(\one-D_\kappa^*)^{-1}(S\,e^{-\kappa})
\;=\;
(e^{2\kappa}-1)^{-1}
\;.
$$
This completes the proof.
\hfill $\Box$

\vspace{.2cm}

Replacing Lemma~\ref{lem-hypsum}, it then follows that 
\begin{eqnarray}
\gamma_\lambda(E) 
\;+\;\imath\,\pi\,\Nn_\lambda(E)
& = & 
\EE_\sigma\;
\frac{1}{L}
\;
\int^E_{-\infty}{de}\;
\partial_e\;
\log
\left(
\,\det\bigl(\,A^e_{\sigma}(\lambda)\,+\,B^e_\sigma(\lambda)\,P_h\,\bigr)\,
\right)
\nonumber
\\
\label{eq-final}
& &
\\
& & \;+\;\frac{2\,\lambda}{L}\;\Re e\;
\Tr\left(
\,
(e^{2\kappa}-\one )^{-1}P_h\;\EE_\sigma(a_\sigma+b_\sigma)\;P_h
\,\right)
\;+\;
\Oo\left(
\frac{\lambda^2}{g_e}\,,\,\frac{\lambda^2}{g_h^3}\right)
\;.
\nonumber
\end{eqnarray}
As this holds also for the translation invariant operator with one fixed
$\sigma$ at every site $n$ (which has non-random independent entries), 
one deduces
$$
\gamma_\lambda(E) 
\;+\;\imath\,\pi\,\Nn_\lambda(E)
\;=\;
\EE_\sigma\;
\gamma_{\lambda,\sigma}(E) 
\;+\;\imath\,\pi\;
\EE_\sigma\;\Nn_{\lambda,\sigma}(E)
\;+\;
\Oo\left(
\frac{\lambda^2}{g_e}\,,\,\frac{\lambda^2}{g_h^3}\right)
\;,
$$
where $\gamma_{\lambda,\sigma}(E)$ and  
$\Nn_{\lambda,\sigma}(E)$ are averaged Lyapunov exponent and IDS of the
translation invariant operator (as already described in the
introduction). This therefore proves Theorem~\ref{theo-perturb}. Expanding the
logarithm in \eqref{eq-final}, some straightforward algebra leads to
more explicit perturbative formulas for
$\gamma_\lambda(E)$ and 
$\Nn_\lambda(E)$ in terms of $\EE_\sigma(\Pp_\sigma)$.
Let us note that the contribution of Lemma~\ref{lem-hypsum} is real and hence
only contributes to $\gamma_\lambda(E)$. The lowest order contribution to
$\Nn_\lambda(E)$ is given by the first term in \eqref{eq-exhelp0} and one can
check that it is only the contribution of $\EE(\Pp_\sigma)$ which changes the
rotation phases of the elliptic channels of $\Rr$.


\end{document}